\documentclass[aps,preprint,tightenlines,superscriptaddress]{revtex4}
\usepackage{epsfig}
\usepackage{graphicx}
\usepackage{psfrag}
\usepackage{amsmath,amssymb}
\usepackage{colordvi}
\usepackage{amsfonts}
\usepackage{enumerate}
\usepackage{slashed}
\usepackage{multirow}
\usepackage{color}

\graphicspath{{ChPT_Graphs/}}

\newcommand{\p}{\partial}
\newcommand{\m}{\mathcal}

\newcommand{\la}{\langle}
\newcommand{\ra}{\rangle}

\newcommand{\tr}{{\rm Tr}}

\newcommand{\ovl}{\overline}

\begin{document}
\title{Chiral Logarithms in $\Delta S=1$ Kaon Decay Amplitudes \\ in General Effective Flavor Theories }
\author{Panying Chen}
\affiliation{Department of Physics, University of Maryland, College
Park, Maryland 20742, USA}
\author{Hongwei Ke}
\affiliation{Department of Physics, University of Maryland, College
Park, Maryland 20742, USA}
\affiliation{Department of Physics, Nankai University, Tianjin, 300071, P. R. China}
\author{Xiangdong Ji}
\affiliation{Department of Physics, University of Maryland, College
Park, Maryland 20742, USA} \affiliation{Center for High-Energy
Physics and Institute of Theoretical Physics, \\ Peking University
Beijing, 100080, P. R. China}

\date{\today}
\vspace{0.5in}
\begin{abstract}
We study the chiral logarithms in $\Delta S=1$ kaon decay amplitudes from new
flavor physics in beyond-standard-model theories. We systematically classify the chiral structures of
dimension-5, 6 and 7 effective QCD operators constructed out of light-quark
(up, down and strange) and gluon fields. Using the standard chiral perturbation theory, we
calculate the leading chiral-logarithms associated with these operators. The result is useful
for lattice calculations of the QCD matrix elements in $K\rightarrow \pi\pi$ decay
necessary, for example, to understand the physical origin of the direct CP violation
parameter $\epsilon'$. As a concrete example, we consider the
new operators present in minimal left-right symmetric models.

\end{abstract}

\maketitle

\section{Introduction}

Non-leptonic kaon decay has been a focus for both theoretical and experimental physics for over 40 years
since the discovery of CP violation by Christenson, Cronin, Fitch and Turlay~\cite{Christenson:1964fg} in  $K_L\rightarrow 2\pi$. Since then the origin of CP violation has long been a challenge to many
theoretical models. The recent data from various experiments have yielded a clear non-vanishing direct CP-violation parameter~\cite{CPV:Exp,PDG}:
\begin{equation}
{\rm Re}(\epsilon'/\epsilon)=(16.7\pm 2.6)\times 10^{-4}\ ,
\end{equation}
which ruled out the so-called superweak theory where no direct CP violation appears in the decay~\cite{Wolfenstein:1964ks}. At present, a full theoretic explanation to the origin of this phenomenon is still lacking. In the framework of the standard model (SM), direct CP violation can be generated by the non-zero phase in the quark flavor-mixing matrix (CKM matrix), as was suggested by Kobayashi and Maskawa~\cite{Kobayashi:1973fv}. A precision calculation of the effect, however, is extremely hard due to the non-perturbative nature of the strong interactions at low energy. Results from several groups utilizing different methods differ widely, with error bars much larger than that of the experimental result~\cite{Buras:2003zz}. The unsatisfying situation of the theoretical calculations have attracted much interest in attributing part of the phenomenon to physics beyond SM.

To be able to pin down the contribution to $\epsilon'$ from models containing new physics, one has to make precision calculations of the strong-interaction physics associated with the non-perturbative structure of kaons and pions. Various methods have been used to calculate the hadronic matrix elements, such as lattice~\cite{Ciuchini:1997kd,Ciuchini:2000zz}, QCD-inspired models~\cite{Bertolini:2000dy,Pallante:2001he}, chiral expansion together with large-$N_c$~\cite{Hambye:1999yy}, and parametrizations~\cite{Bosch:1999wr}.  At present, the lattice field theory is the only approach based on first principles, with controllable systematic errors.  There are difficulties in lattice calculations which are associated with the fact that the final state contains more than one particle. By Maiani-Testa theorem~\cite{Maiani:1990ca}, it is impossible to extract the physical kaon decay matrix elements by taking the limit $\tau\rightarrow\infty$ in the Euclidean space. In practise, there are several ways to avoid it: one can either work with an unphysical choice of momenta~\cite{Bernard:1987pr,Aoki:1997ev}, utilize an unphysical set of meson masses~\cite{Dawson:1997ic,Golterman:1998af}, or derive the physical matrix elements by unphysical, but calculable ones. All of these methods need chiral perturbation theory (ChPT).

ChPT assumes an approximate chiral symmetry exists in $SU(3)_f$ and describes the low-energy QCD physics under a chiral breaking scale $\Lambda_{\chi}\sim m_\rho$ by the pseudo-Goldstone particles, namely pions, kaons and eta. Then the low-energy physics can be perturbatively expanded in powers of the particles' external momenta and masses. It further assumes that, the Wilson coefficients of the QCD operators expanded in terms of meson operators are independent of the external states. Therefore the amplitudes of a large number of reactions can be determined by a relatively small set of coefficients, which gives us the predicting power. In the case of kaon decay, ChPT is used to connect the desired matrix element $\langle\pi\pi|\mathcal{O}|K\rangle$ with some unphysical quantities, such as $\langle\pi|\mathcal{O}|K\rangle$ and $\langle 0|\mathcal{O}|K\rangle$. The results in ChPT are needed before doing relevant lattice calculations. Here we will neglect some subtleties in the ChPT (such as quadratic divergence cancelations, zero pion mass corrections, etc) and focus on possible operator structures as well as their chiral logarithm corrections for the kaon decay process.  It is the goal of this paper to examine the chiral structures of possible QCD operators responsible for $\Delta s=1,\Delta d=-1$ decay in generic beyong-SM theories and to calculate the large chiral logarithms associated with them. Previous calculations have been made for operators present in the SM~\cite{Bijnens:1984ec,Bijnens:1984qt,Golterman:1997st,Golterman:2000fw,Bijnens:1998mb,Cirigliano:1999pv}. Our work extends these studies to all possible operators in new physics models.

The paper is organized as follows: we start from the operator basis in SM for the kaon decay, as well as possible new operators coming from physics beyond SM. A chiral perturbation theory calculation will be presented in the following section, with all corresponding operators and their one-loop corrections of the matrix elements. We end this section by applying our result to a specific example. Concluding remarks and outlook are presented in the last section.

\section{Effective Operators from New Flavor Physics}

In this section, we consider effective QCD operators contributing to CP-violating $K\rightarrow \pi\pi$ decay in a generic weak-interaction theory.  There is an extensive literature on this topic in the context of SM~\cite{Bertolini:1998vd,Buras:1998raa}. Our focus is on new operators arising from novel CP-violating mechanisms beyond SM.  We classify the effective operators in terms of their flavor symmetry properties under chiral group $SU(3)_L\times SU(3)_R$ when up, down, and strange quarks are taken as light.

The direct CP violation parameter $\epsilon'$ for $K\rightarrow \pi\pi$ decay is defined as~\cite{Bertolini:1998vd}:
\begin{equation}
\epsilon'=\frac{1}{\sqrt{2}}e^{\left(\frac{\pi}{2}+\delta_2-\delta_0\right)}\frac{{\rm Re}A_2}{{\rm Re}A_0} \left(\frac{\mbox{Im}A_0}{\mbox{Re}A_0}-\frac{\mbox{Im}A_2}{\mbox{Re}A_2}\right) \ ,
\end{equation}
where $\delta_I$ is the strong-interaction $\pi\pi$ scattering phase shifts, and $A_I$ is the weak kaon decay amplitudes:
\begin{equation}
A_I e^{i\delta_I}=\langle\pi\pi(I=0,2)|(-i\mathcal{H}_W)|K^0\rangle \ ,
\end{equation}
where $\mathcal{H}_W$ is the effective weak-interaction hamiltonian which depends on the underlying theory of kaon decay.
The small ratio $\omega\equiv{\rm Re}A_2/{\rm Re}A_0\approx1/22$ reflects the well-known $\Delta I=1/2$ rule. Accurate calculations of $\epsilon'$ depend on reliable evaluations of the effective QCD operators present in ${\cal H}_W$. Our goal in this paper is to
classify these QCD operators and study their chiral behavior.

\subsection{Standard Model Operators}

The standard procedure for calculating $\epsilon'$ utilizes an effective field theory approach.
The physics at high energy (or short distance) can be calculated perturbatively and is included in Wilson coefficients.
The physics at low-energy scales is included in the effective QCD operators composed of light flavor quark fields ($u,d,s$) and
gluon fields. Large QCD radiative corrections or large logarithms are resumed by solving renormalization group equations.
The effective operators responsible for the neutral kaon decay have the flavor quantum numbers $\Delta s=1,\,\Delta d=-1$.
In the SM, it is well-known that $\cal H^{\rm eff}$ consists of following 10 operators~\cite{Buchalla:1995vs, Buras:1998raa}:
\begin{eqnarray}
\label{smop}
Q_1&=&(\overline{s}_i  u_j )_{V-A}(\overline{u}_j  d_i )_{V-A}  \ , \nonumber\\
Q_2&=&(\overline{s}_i  u_i )_{V-A}(\overline{u}_j  d_j )_{V-A} \ , \nonumber\\
Q_{3,5}&=&(\overline{s}_i  d_i )_{V-A}\sum_q(\overline{q}_j  q_j )_{V\mp A}\ , \nonumber\\
Q_{4,6}&=&(\overline{s}_i  d_j )_{V-A}\sum_q(\overline{q}_j  q_i )_{V\mp A}\ , \nonumber\\
Q_{7,9}&=&\frac{3}{2}(\overline{s}_i  d_i )_{V-A}\sum_q e_q(\overline{q}_j  q_j )_{V\pm A}\ , \nonumber\\
Q_{8,10}&=&\frac{3}{2}(\overline{s}_i  d_j )_{V-A}\sum_q e_q(\overline{q}_j  q_i )_{V\pm A} \ ,
\end{eqnarray}
where $(\bar qq')_{V\pm A} = \bar q_{L,R} \gamma_\mu q_{L,R}$ with $q_{L,R}$ representing the left-(right-) handed quark fields. The summation in $q$ is over the light-quark flavors; $u,d,s$; $i$ and $j$ are color indices;
and $e_q$ is the algebraic charge factor for flavor $q$. The $Q_{1,2}$ come from the single $W_L$-boson exchange tree diagram, and $Q_{3-6},\;Q_{7-10}$ are derived from one-loop gluon and electro-weak penguin diagrams, respectively. In the SM, $({\rm Im}A_0/{\rm Re}A_0)$ is dominated by the QCD penguin operators, whereas $({\rm Im}A_2/{\rm Re}A_2)$ receives contribution from the electro-weak penguin operators only, because the gluon interaction is flavor-singlet and cannot contribute in $\Delta I=3/2$ channel.

A chiral structure analysis of the above ten operators will be useful if we wish to use lattice QCD to calculate the relevant matrix elements in kaon decays. Using $(m,n)$ to denote a representation of group $SU_L$(3)$\times$ $SU_R$(3), where $m$ and $n$ are the dimensions of SU(3) representations, it is then easy to see that $Q_{1,2}$ and $Q_{9,10}$ belong to $(8,1)$ and $(27,1)$, $Q_{3\sim6}$ to $(8,1)$, and $Q_{7,8}$ to $(8,8)$~\cite{Buras:1998raa}. ChPT calculations have been made to uncover the large logarithms associated with these operators, which in turn help to establish relations of different matrix elements useful for lattice QCD calculations. For the reader's convenience, we have collected
the standard chiral results in the Appendix.

The question is what is the general chiral structure of all possible weak operators that
might emerge in theories beyond SM? The rest of this section is devoted
to addressing this question.

\subsection{Dimension-5 and -6 Operators}

Let us systematically consider the possible operators and their chiral structures in a general low-energy
description of kaon decay, independent of the underlying short-distance flavor physics that
can be taken into account by Wilson coefficients. The lowest dimensional operator is a dimension-5
chromo-magnetic operator,
\begin{equation}
Q_M =  \bar s (\sigma^{\mu\nu})t^a d ~G^a_{\mu\nu} \ .
\end{equation}
This operator does appear in the standard model through penguin diagram as shown in left-panel Fig. 1, although
it is proportional to the strange or down quark masses, which is chirally suppressed. It also appears
naturally in the left-right symmetric model (LRSM) with left-right handed gauge-boson mixing,
proportional to charm or top quark masses~\cite{Ecker:1985vv}, as shown on the right panel
in Fig.1. Under chiral symmetry, this operator transforms
as $(\ovl 3,3) + (3, \ovl 3)$. The chiral logarithms appearing with the matrix elements
of this operator has been studied before in the literature, and is collected in the Appendix.

\begin{figure}[!htbp]
\begin{center}
\includegraphics[scale=0.8]{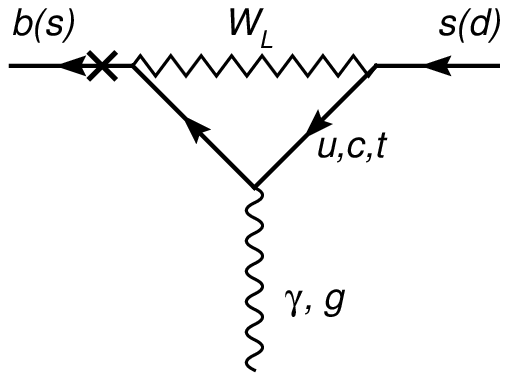}
\hspace{1cm}
\includegraphics[scale=0.8]{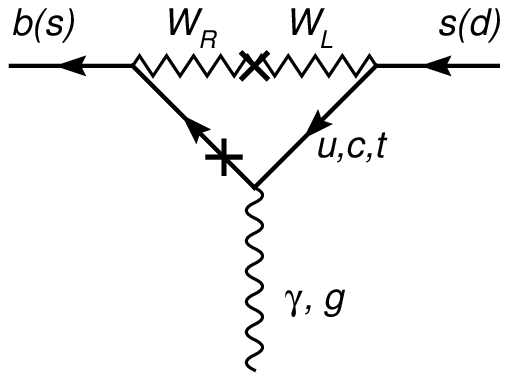}
\caption{Dimension-5 effective operators generated from the weak-interaction vertex corrections
in SM (left) and in LRSM (right). The crosses on fermion lines represent mass insertion, needed to flip the chirality of the quarks.}
\label{B2S}
\end{center}
\end{figure}

\begin{figure}[!htbp]
\begin{center}
\includegraphics[scale=0.8]{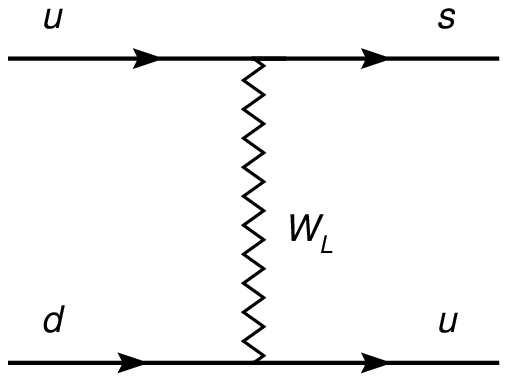}
\hspace{1cm}
\includegraphics[scale=0.8]{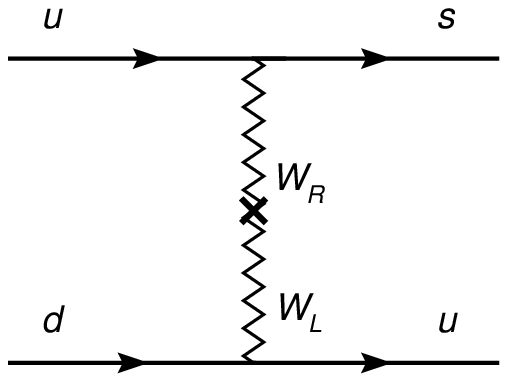}
\caption{Feynman diagrams generating dimension-6 quark operators in SM and LRSM.}
\label{WLWR}
\end{center}
\end{figure}

Next, consider dimension-6 four-quark operators. We define the following flavor tensor
\begin{equation}
                \Theta^{jl}_{ik} =     (\bar q^i \Gamma q_j)(\bar q^k \Gamma' q_l) \ ,
\end{equation}
where the flavor indices $i,j,k,l$ go through 1, 2, and 3, or up, down and strange quarks.
 $\Gamma$ and $\Gamma'$ are possible Dirac matrix structures. In addition, there are
two independent color structures $({\bf 1})({\bf 1})$ and $(t^a)(t^a)$ which are not
essential for the following discussion. Assuming that all fields are projected to their
helicity states, the possible helicities are as follows:
\begin{itemize}
\item{ All four quark fields have the same chiral projection}
\item{ Both $\bar q^i$ and $\bar q^k$ (also $j$ and $l$) have the opposite chiral projection}
\item{ Both $\bar q^i$ and $\bar q^k$ (also $j$ and $l$) have the same chiral projection}
\end{itemize}
These are only possibilities because the operators must be Lorentz scalars, and the numbers of
left and right-handed fields apart from the first case must be exactly 2, respectively.
In the first and second case, the operators are the ones appearing in the SM weak interactions,
as shown in Eq. (\ref{smop}), and their parity partners. They correspond to chiral structures (8,1), (1,8), (27, 1), and (1,27) from the first case, and
(8,1), (1,8) and (8,8) from the second case. The new (1,8) and (1,27) structures will appear in,
for example, LR symmetric models where the right-handed gauge boson plays the same role
as left-handed one in the SM.
Since the strong interactions conserve parity, the new operators in LRSM have the same matrix elements
as $Q_i$'s in SM up to a parity sign.  The corresponding Feynman diagrams in both SM and LRSM for the first and second cases are shown in Fig. \ref{WLWR}.

In the last case, there are new chiral structures arising from operators of type
\begin{equation}
                    (\bar s_L \Gamma d_R)(\bar q_L \Gamma q_R), ~~~  (\bar s_R \Gamma d_L)(\bar q_R \Gamma q_L)\ .
\end{equation}
where two $\Gamma's$ must be the same. The new chiral structures are $(\ovl 6, 6)$, $(\ovl 6, \ovl 3)$, $(3, 6)$,
and their parity conjugates. However, $(\ovl{6},\ovl{3})$ and $(3,6)$
involve symmetrization of two flavor indices and, at the same time, anti-symmetrization of the other two. It is easy to check that
the result vanishes, and we are left with just $(\ovl 6, 6)$ and $(6, \ovl 6)$.

Let us consider the following tensor with up (and hence lower) indices symmetrized
\begin{equation}
                  \Theta^{kl}_{(ij)} = \frac{1}{2}(\bar q_L^i \Gamma q_{Rk}) (\bar q_L^j \Gamma q_{Rl})
                    + \frac{1}{2}(\bar q_L^j \Gamma q_{Rk}) (\bar q_L^i \Gamma q_{Rl}) \ ,
\end{equation}
where the upper indices represents the left-handed fields and the lower indices the right-handed.
Without loss of generality, we take $i=3$. If $j=3$, and $k$ or $l$ is $3$ and the other indices must be a 2,
one get an isospin-1/2 operator
\begin{equation}
                \Theta^{(\ovl{6},6)}_{1/2,A}\equiv\Theta^{23}_{33} = \bar s_L \Gamma d_R \bar s_L \Gamma s_R \ .
\end{equation}
If we define a tensor $T^{ij}_{kl}$ which multiplies the quark operator $\Theta^{kl}_{ij}$ to generate the above
operator, $T^{ij}_{kl}\Theta^{kl}_{ij}$, we have
\begin{equation}
                T^{33}_{23}  = T^{33}_{32} = 1/2 \ ,
\end{equation}
and other components zero.

\begin{figure}[!htbp]
\begin{center}
\includegraphics[scale=0.8]{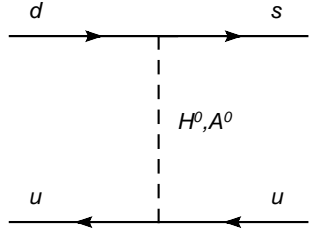}
\caption{Feynman diagrams generating scalar quark interactions through neutral-current Higgs exchanges.}
\label{Higgs}
\end{center}
\end{figure}

On the other hand, if $j$, $k$, $l$ take 1's and 2's, one can subtract the trace with respect to $j$
and $k$, and $j$ and $l$, and one obtain an isospin-3/2 operator
\begin{eqnarray}
\Theta^{(\ovl{6},6)}_{3/2} &\equiv&  \Theta_{(31)}^{12}  + \Theta_{(31)}^{21} - \Theta_{(32)}^{22} \nonumber \\
                           &   =  & \bar s_L\Gamma u_R \bar u_L \Gamma d_R + \bar s_L\Gamma d_R \bar u_L \Gamma u_R
                               - \bar s_L \Gamma d_R \bar d_L \Gamma d_R \ ,
\end{eqnarray}
with corresponding non-zero tensor components
\begin{equation}
             T^{31}_{12} = T^{31}_{21} = T^{13}_{12} = T^{13}_{21} = - T^{32}_{22} = - T^{23}_{22} = 1/2 \ .
\end{equation}
Another isospin-1/2 operator will can be obtained its trace part,
\begin{eqnarray}
\Theta^{(\ovl{6},6)}_{1/2,S} & \equiv&  \Theta_{(31)}^{12}  + \Theta_{(31)}^{21} +2 \Theta_{(32)}^{22} \nonumber \\
                             &   =   & \bar s_L\Gamma u_R \bar u_L \Gamma d_R + \bar s_L\Gamma d_R \bar u_L \Gamma u_R
                                    +2 \bar s_L \Gamma d_R \bar d_L \Gamma d_R \ ,
\end{eqnarray}
and the corresponding tensor components are,
\begin{equation}
             T^{31}_{12} = T^{31}_{21} = T^{13}_{12} = T^{13}_{21} = 1/2; ~~~ T^{32}_{22} = T^{23}_{22} = 1 \ .
\end{equation}
An example of these new operators in LRSM through flavor-changing neutral and charged currents is
shown in Fig. \ref{Higgs}. The relevant QCD four-quark operator will be:
\begin{eqnarray}
\mathcal{O}^{\Delta s=1} &=& (\bar{s}_L d_R)\sum_q(\bar{q}_L q_R) \nonumber \\
 &=& \frac{1}{2}\left[\Theta^{(\ovl{6},6)}_{1/2,S}+2\Theta^{(\ovl{6},6)}_{1/2,A}-\Theta^{(3,\ovl{3})}\right]
\end{eqnarray}
where $\Theta^{(3,\ovl{3})}\equiv(\bar{s}_L u_R)(\bar{u}_L d_R)-(\bar{s}_L d_R)(\bar{u}_L u_R)$.

\subsection{Dimension-7 Operators}

Dimension-7 operators come in two types. The first is the chromomagnetic operators with an
insertion of two additional derivatives, which does not change the original chiral structure.
The second type is an insertion of one derivative into four-quark operators discussed above.
Since the covariant derivative has one Lorentz index, it must be contracted with
another Lorentz index appearing on a Dirac matrix. An example of this type of operators is
\begin{eqnarray}
\mathcal{O}^{LR}_{P,g} &=& \bar s_{i,L} (i\sigma^{\mu\nu}) d_{j,R}\sum_q \bar q_{j,L} (\gamma_\mu D_\nu) q_{i,L} \ , \nonumber\\
\mathcal{O}^{LR}_{P,EW} &=& \bar s_{i,L} (i\sigma^{\mu\nu}) d_{i,R}\sum_q e_q\bar q_{j,L} (\gamma_\mu D_\nu) q_{j,L} \ ,
\end{eqnarray}
which comes from the gluon and electromagnetic penguin diagram in LRSM as shown in Fig. \ref{B2S}.
These operators contain either 3 left-handed fields and 1 right-handed one, or 3 right-handed fields and 1 left handed one.
They have novel chiral structures $(15,\ovl{3})$, $(\ovl{15}, 3)$, $(6,3)$, $(\ovl 6,\ovl 3)$, and parity  partners. Let us classify them all in details.

\subsubsection{$(\ovl{15},3)$}
Let us use $\Theta_{ij}^{kl}$ to represent an operator $\bar q_L^i\Gamma q_{Lk} \bar q_L^j \Gamma ' q_{Rl}$, where $l$ is flavor index of the right-handed field, and $\Gamma$ and $\Gamma'$ are
not just Dirac matrices.
We first construct $\hat \Theta_{ij}^{kl}$ which forms $(\ovl{15},3)$
after symmetrizing the up two indices and subtracting the traces,
\begin{equation}
      2\hat \Theta_{ij}^{kl} =  \Theta_{ij}^{kl} + \Theta_{ji}^{kl} - \frac{1}{4}\delta_i^k
        \left[\Theta_{\alpha j}^{\alpha l} + \Theta_{ j\alpha}^{\alpha l}\right] - \frac{1}{4}
          \delta^k_j  \left[\Theta_{\alpha i}^{\alpha l} + \Theta_{ i\alpha}^{\alpha l}\right] \ ,
\label{(15b,3)}
\end{equation}
where $\alpha$ sums over 1, 2, and 3.
Clearly, $i$, or equivalently $j$, has to be an $\bar s$. One isospin-1/2 operator that one can immediately identity is when $j$ is 3, $k$ is 2, and $l$ is 3, namely
\begin{equation}
\Theta^{(\ovl{15},3)}_{1/2}\equiv\hat \Theta_{33}^{23} = \Theta_{33}^{23}  =
                     \bar s_L \Gamma d_L \bar s_L\Gamma's_R \ ,
\end{equation}
where the only non-zero tensor component is
\begin{equation}
                T^{33}_{23}  = 1 \ .
\end{equation}
Other independent operators can be obtained by considering $j$, $k$, and $l$ as up and down quarks.
Others, such as $\hat \Theta^{32}_{33}$, can be related to these through traceless conditions.

One can get an isospin-3/2 operator by symmetrizing $k$ and $l$ while taking away SU(2) traces between $j$ and $k$, and $j$ and $l$.
Thus we have the following combination,
\begin{equation}
    \hat \Theta_{3j}^{kl} + \hat \Theta_{3j}^{lk} - \frac{1}{3}\delta^k_j \left(  \hat \Theta_{3a}^{al} + \hat \Theta_{3a}^{la}\right)
       - \frac{1}{3}\delta_j^l\left(  \hat \Theta_{3a}^{ak} + \hat \Theta_{3a}^{ka}\right) \ ,
\end{equation}
where $a$ sums over 1 and 2 only, and $j$, $k$ and $l$ can take value in 1 or 2. There is only one independent operator
\begin{eqnarray}
\Theta^{(\ovl{15},3)}_{3/2}  &\equiv&  2\left(\hat \Theta_{31}^{21} + \hat \Theta_{31}^{12} -\hat \Theta_{32}^{22}\right) \nonumber\\
  &=& \bar s_L \Gamma u_L \bar u_L \Gamma' d_R + \bar s_L \Gamma d_L \bar u_L \Gamma' u_R
  + \bar u_L \Gamma u_L \bar s_L \Gamma' d_R \nonumber \\ && + \bar u_L \Gamma d_L \bar s_L \Gamma' u_R
  - \bar s_L \Gamma d_L \bar d_L \Gamma' d_R - \bar d_L \Gamma d_L \bar s_L \Gamma' d_R \ .
\end{eqnarray}
The corresponding tensor components are
\begin{equation}
      T^{31}_{12} = T^{13}_{12} = T^{31}_{21} = T^{13}_{21}= - T^{32}_{22} =- T^{23}_{22}=1 \ .
\end{equation}
The trace part of the above operator produces an $I=1/2$ operator, $ \hat \Theta_{3a}^{a2} + \hat \Theta_{3a}^{2a} $.  We can subtract from the result with another isospin-1/2 operator, $\hat \Theta^{33}_{32}$, to cancel the unwanted trace part $-\frac{1}{4}\delta_i^{k}[\Theta_{\alpha j}^{\alpha l}]-\frac{1}{4}\delta^k_{j}[\Theta_{\alpha i}^{\alpha l}]$ in Eq. (\ref{(15b,3)}).
The resulting $I=1/2$ operator is
\begin{eqnarray}
\Theta^{(\ovl{15},3)}_{1/2,S}  &\equiv& 2\left(\hat \Theta_{31}^{12} + \hat \Theta_{31}^{21} + 2\hat \Theta_{32}^{22}\right) - 3\hat \Theta_{33}^{32}\nonumber \\
  & = &   \bar s_L \Gamma u_L \bar u_L \Gamma' d_R + \bar u_L \Gamma u_L \bar s_L \Gamma' d_R
  + \bar s_L \Gamma d_L \bar u_L \Gamma' u_R \nonumber \\ && + \bar u_L \Gamma d_L \bar s_L \Gamma' u_R
  + 2\bar s_L \Gamma d_L \bar d_L \Gamma' d_R + 2\bar d_L \Gamma d_L \bar s_L \Gamma' d_R - 3\bar s_L \Gamma s_L \bar s_L \Gamma' d_R  \ ,
\end{eqnarray}
with the following tensor components,
\begin{equation}
   T^{31}_{21} = T^{13}_{21} = T^{31}_{12} = T^{13}_{12}= \frac{1}{2} T^{32}_{22} = \frac{1}{2} T^{23}_{22}= \frac{1}{3} T^{33}_{32}=1 \ .
\end{equation}

Finally, one can antisymmetrize $k$ and $l$ to generator another isospin 1/2 operator.  Again we add $\hat \Theta_{33}^{32}$ to cancel the unwanted trace part:
\begin{eqnarray}
\Theta^{(\ovl{15},3)}_{1/2,A} &\equiv& 2\left(\hat \Theta_{31}^{21} - \hat \Theta_{31}^{12}\right)+ \hat \Theta_{33}^{32} \nonumber \\
                    & = & \bar s_L \Gamma d_L \bar u_L \Gamma' u_R + \bar u_L \Gamma d_L \bar s_L \Gamma' u_R
  - \bar s_L \Gamma u_L \bar u_L \Gamma' d_R \nonumber \\ && - \bar u_L \Gamma u_L \bar s_L \Gamma' d_R
  + \bar s_L \Gamma s_L \bar s_L \Gamma' d_R \ ,
\end{eqnarray}
with the following tensor components
\begin{equation}
 T^{31}_{21} = T^{13}_{21} = -T^{31}_{12} = -T^{13}_{12} = T^{33}_{32} = 1 \ .
\end{equation}
Note that the operators in  $(3, \ovl{15})$ can be obtained from the above
through parity transformation.

\subsubsection{$(\ovl{3}, 15)$}

Define $\Theta_{ij}^{kl} = \bar q_L^i \Gamma q_{Rk} \bar q_R^j \Gamma' q_{Rl}$, and construct the general
$(\ovl{3}, 15)$ operators,
\begin{equation}
      2\hat \Theta_{ij}^{kl} =  \Theta_{ij}^{kl} + \Theta_{ij}^{lk} - \frac{1}{4}\delta^k_j
        \left[\Theta_{i\alpha }^{\alpha l} + \Theta_{i\alpha}^{l\alpha}\right] - \frac{1}{4}
         \delta_j^l  \left[\Theta_{i\alpha }^{\alpha k} + \Theta_{ i\alpha}^{k\alpha }\right] \ ,
\end{equation}
where the $\alpha$ trace is over 1, 2, and 3. Either index $i$ or $j$ can be identified as the strange quark
field. In either case, one can construct isospin-3/2 operators by subtracting SU(2) trace.
With the left-handed strange quark, we have
\begin{eqnarray}
\Theta^{(\ovl{3},15)}_{3/2,L} &\equiv&  \hat \Theta_{31}^{12} + \hat \Theta_{31}^{21} - \hat \Theta_{32}^{22}
      =  \Theta_{31}^{12} +  \Theta_{31}^{21} -  \Theta_{32}^{22}  \nonumber \\
      &=& \bar s_L \Gamma u_R \bar u_R \Gamma' d_R +
         \bar s_L \Gamma d_R \bar u_R \Gamma' u_R  - \bar s_L \Gamma d_R \bar d_R \Gamma' d_R \ ,
\end{eqnarray}
with tensor components,
\begin{equation}
     T^{31}_{12} = T^{31}_{21} = - T^{32}_{22} = 1 \ .
\end{equation}
With the right-handed strange quark,
\begin{eqnarray}
\Theta^{(\ovl{3},15)}_{3/2,R} &\equiv& \hat \Theta_{13}^{12} + \hat \Theta_{13}^{21} - \hat \Theta_{23}^{22}
     =  \Theta_{13}^{12} +  \Theta_{13}^{21} -  \Theta_{23}^{22}  \nonumber \\
      &=& \bar u_L \Gamma u_R \bar s_R \Gamma' d_R +
         \bar u_L \Gamma d_R \bar s_R \Gamma' u_R  - \bar d_L \Gamma d_R \bar s_R \Gamma' d_R \ ,
\end{eqnarray}
with tensor components,
\begin{equation}
     T^{13}_{12} = T^{13}_{21} = - T^{23}_{22} = 1 \ .
\end{equation}

There are also two isospin-1/2 operators. The first one with left-handed strange quark,
\begin{eqnarray}
\Theta^{(\ovl{3},15)}_{1/2,L}&\equiv&4\left(\hat \Theta_{31}^{12} + \hat \Theta_{31}^{21} +2 \hat \Theta_{32}^{22} \right)\nonumber \\
      &=& \bar s_L \Gamma u_R \bar u_R\Gamma' d_R + \bar s_L \Gamma d_R \bar u_R\Gamma' u_R
        + 2\bar s_L \Gamma d_R \bar d_R\Gamma' d_R
        \nonumber \\  && - 3 \bar s_L \Gamma s_R \bar s_R\Gamma' d_R
          - 3 \bar s_L \Gamma d_R \bar s_R\Gamma' s_R
\end{eqnarray}
with tensor components
\begin{equation}
   T^{31}_{12} = T^{31}_{21} = \frac{1}{2}T^{32}_{22} = -\frac{1}{3}T^{33}_{32} = -\frac{1}{3}T^{33}_{23}= 1 \ .
\end{equation}
And the second one has the right-handed strange quark,
\begin{eqnarray}
\Theta^{(\ovl{3},15)}_{1/2,R} &\equiv& \hat \Theta_{13}^{12} + \hat \Theta_{13}^{21} +2 \hat \Theta_{23}^{22} =
 \Theta_{13}^{12} + \Theta_{13}^{21} +2  \Theta_{23}^{22} \nonumber \\
  &=& \bar u_L \Gamma u_R \bar s_R \Gamma' d_R +  \bar u_L \Gamma d_R \bar s_R \Gamma' u_R
   + 2  \bar d_L \Gamma d_R \bar s_R \Gamma' d_R
\end{eqnarray}
with tensor components
\begin{equation}
   T^{13}_{12} = T^{12}_{21} = \frac{1}{2}T^{23}_{22} = 1 \ .
\end{equation}
The $\hat \Theta_{33}^{32}$ is not independent by the same reason as for $(\ovl{15},3)$.

\subsubsection{$(6, 3)$}

Define $\Theta_{ij}^{kl} = \bar q^i_L\Gamma q_{Lk} \bar q^j_L \Gamma' q_{Rl}$, and
construct the $(6,3)$ operator,
\begin{equation}
  \hat \Theta_{ij}^{kl}=\epsilon_{ijm}\hat \Theta^{(mk)l} = \frac{1}{2}\epsilon_{ijm}\left(\epsilon^{\alpha\beta m} \Theta_{\alpha\beta}^{kl} + \epsilon^{\alpha\beta k} \Theta_{\alpha\beta}^{ml}\right) \ .
\end{equation}
where $k$, $m$ are symmetric, and $\alpha$, $\beta$ run over 1 to 3.  To get the $\Delta s= -\Delta d=1$ operators, none of the $m$, $k$ and $l$ can be a 3: when 3 is
on the $\epsilon$, it prevents
both $i$ and $j$ from being a strange quark, and when 3 is a lower index, both $i$ and $j$
must be 3 which is impossible because of the antisymmetry. In fact, the only possible combination for
$m$, $k$ and $l$ is 2, 2, 1.

To get an isospin 3/2 operator, one must symmetrize $m$, $k$ and $l$, yielding
\begin{eqnarray}
\Theta^{(6,3)}_{3/2}  &\equiv& \hat \Theta^{221}+\hat \Theta^{212}+\hat \Theta^{122} = \hat \Theta_{31}^{21} + \hat \Theta_{31}^{12} +\hat \Theta_{23}^{22} \nonumber\\
  &=& \bar s_L \Gamma d_L \bar u_L \Gamma' u_R + \bar s_L \Gamma u_L \bar u_L \Gamma' d_R
  - \bar u_L \Gamma d_L \bar s_L \Gamma' u_R \nonumber \\ && - \bar u_L \Gamma u_L \bar s_L \Gamma' d_R
  - \bar s_L \Gamma d_L \bar d_L \Gamma' d_R + \bar d_L \Gamma d_L \bar s_L \Gamma' d_R \ .
\end{eqnarray}
The corresponding tensor components are
\begin{equation}
      T^{31}_{12} = -T^{13}_{12} = T^{31}_{21} = -T^{13}_{21}= - T^{32}_{22} = T^{23}_{22}=1 \ .
\end{equation}
There is also an isospin-1/2 operator by anti-symmetrizing $k$ and $l$
\begin{eqnarray}
\Theta^{(6,3)}_{1/2}  &\equiv&  2 \hat \Theta^{221} - \hat \Theta^{212} - \hat \Theta^{122} = 2\left(\hat \Theta_{31}^{21} - \hat \Theta_{31}^{12}\right)\nonumber\\
  &=& 2\bar s_L \Gamma d_L \bar u_L \Gamma' u_R - 2\bar u_L \Gamma d_L \bar s_L \Gamma' u_R
  - \bar s_L \Gamma u_L \bar u_L \Gamma' d_R \nonumber \\ && + \bar u_L \Gamma u_L \bar s_L \Gamma' d_R
  - \bar d_L \Gamma d_L \bar s_L \Gamma' d_R + \bar s_L \Gamma d_L \bar d_L \Gamma' d_R \ ,
\end{eqnarray}
The tensor components are
\begin{equation}
      \frac{1}{2}T^{31}_{21} = -\frac{1}{2}T^{13}_{21} = -T^{31}_{12} = T^{13}_{12} =  T^{32}_{22} = - T^{23}_{22}=1 \ .
\end{equation}

\subsubsection{$(\ovl{3}, \ovl{6})$}
Define $\Theta_{ij}^{kl} = \bar q^i_L\Gamma q_{Rk} \bar q^j_R \Gamma' q_{Rl}$, and
construct the $(\ovl{3},\ovl{6})$ operator,
\begin{equation}
  \hat \Theta_{ij}^{kl}= \epsilon^{klm}\hat \Theta_{i(jm)} = \frac{1}{2}\epsilon^{klm}\left(\epsilon_{\alpha\beta m} \Theta_{ij}^{\alpha\beta} + \epsilon_{\alpha\beta j} \Theta_{im}^{\alpha\beta}\right) \ ,
\end{equation}
where $j$, $m$ are symmetric, and $\alpha$, $\beta$ run from 1 to 3.
The only choices which will generate the $\Delta s = -\Delta d = 1$ operators is
$i$, $j$ and $m$ takes $3, 3, 1$. If $i$ is 3, and $j$ and $m$ take 3 and 1,
we have the isospin 1/2 operator
\begin{eqnarray}
\Theta^{(\ovl{3},\ovl{6})}_{1/2,L}  &\equiv& \hat \Theta_{313} +\hat \Theta_{331}=2\hat \Theta_{31}^{12} \nonumber \\
               &=& \bar s_L \Gamma u_R \bar u_R \Gamma' d_R -  \bar s_L \Gamma d_R \bar u_R \Gamma' u_R
               + \bar s_L \Gamma d_R \bar s_R \Gamma' s_R - \bar s_L \Gamma s_R \bar s_R \Gamma' d_R \ ,
\end{eqnarray}
with the following tensor components,
\begin{equation}
          T^{31}_{12} = - T^{31}_{21} = T^{33}_{23} = - T^{33}_{32} = 1 \ .
\end{equation}
On the other hand, if $i$ takes 1, one gets another isospin 1/2 operator
\begin{eqnarray}
\Theta^{(\ovl{3},\ovl{6})}_{1/2,R} & \equiv & \hat \Theta_{133} = \hat \Theta_{13}^{12} \nonumber\\
    & = & \bar u_L \Gamma u_R \bar s_R \Gamma' d_R - \bar u_L \Gamma d_R \bar s_R \Gamma' u_R  \ ,
\end{eqnarray}
with the following tensor components
\begin{equation}
       T^{13}_{12} = - T^{13}_{21} = 1 \ .
\end{equation}

One could consider operators with dimension 8 and higher. However, generally they are suppressed
by $1/\Lambda^2$ relative to those
we have considered, where $\Lambda$ is some weak interaction scale.
We summarize the above result in the following table:

\begin{table}[!htb]
\caption{Chiral representations appearing in dimension-5, 6 and 7 operators.}
\begin{tabular}
{|c||c|c|c|}\hline
\label{table}
 & dimension-5 & dimension-6 & dimension-7 \\
 \hline
 &  & (8,1),(1,8) &  \\
 &  &\;\;$(27,1),(1,27)$\;\; & \;\;$(15,\ovl 3),(\ovl 3,15),(\ovl {15},3),(3,\ovl {15})$\;\;\\
 \;\;$(L,R)$\;\;& \;\;$(3,\ovl 3),(\ovl 3,3)$\;\;& (8,8) &$(\ovl 6,\ovl 3),(\ovl 3,\ovl 6),(6,3),(3,6)$ \\
 &  & $(6,\ovl 6),(\ovl 6, 6)$ & $(3,\ovl 3),(\ovl 3,3)$\\
 &  & $(3,\ovl 3),(\ovl 3,3)$ & \\
\hline
\end{tabular}
\end{table}
We emphasize that these operators are completely general, independent of the underlying mechanisms (supersymmetry, large-extra dimension, or little Higgs, etc) for flavor and CP-violations in beyond SM theories.

\section{Chiral Expansion at Leading Order}

Chiral perturbation theory (ChPT) for kaon decay is useful for two reasons: First, it allows one to
connect the physical matrix elements $\la K|\mathcal{O}|\pi\pi\ra$
to some unphysical, but easier-to-calculate matrix elements on lattice.
Second, it yields dependence of the matrix elements on meson masse parameters.
Since lattice calculations are usually done at larger and unphysical meson masses
because of limited computational resources, this dependence can be used to extrapolate the
calculated matrix elements to physical ones. In this section, we will
build a set of effective operators in ChPT up to the lowest order, and use
an example to illustrate how to connect the unphysical processes to
the physical process $K\rightarrow\pi\pi$ we are interested in.

In lattice calculations the quenched approximation to QCD has usually been applied
in the past, where valance quark fields are ``quenched'' by corresponding ghost
quark fields with the same masses and quantum numbers but opposite statistics.
The ChPT can be adapted with the quenched QCD by introducing the ``super-$\eta'$''
field into the effective lagrangian~\cite{Bernard:1992mk}. In this paper we will work
with the full dynamical QCD only.

\subsection{ChPT and SM Operators}

The standard ChPT starts with the nonlinear Goldstone meson field $\Sigma$ by:
\begin{equation}
\Sigma\equiv\exp\left(\frac{2i\phi}{f}\right) \ ,
\end{equation}
where $\phi$ is the Goldstone meson matrix
\begin{eqnarray}
\phi=\left(
\begin{array}{ccc}
\frac{\pi^0}{\sqrt{2}}+\frac{\eta}{\sqrt{6}} & \pi^+ & K^+\\
\pi^- & -\frac{\pi^0}{\sqrt{2}}+\frac{\eta}{\sqrt{6}} & K^0\\
K^- & \ovl{K^0} & -\sqrt{\frac{2}{3}}\eta
\end{array}\right),
\end{eqnarray}
and $f\approx135~{\rm MeV}$ is the bare pion decay constant.  We separate the effective ChPT Lagrangian into two parts:
\begin{equation}
\mathcal{L}_{\rm ChPT}=\mathcal{L}_s+\mathcal{L}_w  \ ,
\end{equation}
where $\mathcal{L}_s$ corresponds the QCD strong interaction which preserves
the flavor symmetry; the $\mathcal{L}_w$ is an effective Lagrangian for non-leptonic
weak interaction, and is responsible for the $\Delta s=1$ processes. The lowest-order
terms for the strong interaction part is:
\begin{equation}
\mathcal{L}_{s}^{(2)}=\frac{f^2}{8}\tr(\p_\mu\Sigma\p^\mu\Sigma)+v\tr\left[M\Sigma+(M\Sigma)^\dagger\right],
\end{equation}
where $M\equiv{\rm diag}(m_u,m_d,m_s)$ is the quark mass matrix; and $v\sim-\frac{1}{2}\la\bar{u}u\ra$ is proportional to the quark chiral condensate at chiral limit.  We demand the fields transform under $SU(3)_L\times SU(3)_R$ as:
\begin{equation}
\Sigma\rightarrow L \Sigma R^\dagger,\;\;\;M \rightarrow R \Sigma L^\dagger,
\end{equation}
to keep the Lagrangian invariant under an $SU(3)_L\times SU(3)_R$ transformation.
Higher order terms in the effective Lagrangian contain higher derivatives, and can be
written in systematic derivative and mass expansion. For our purpose here, however,
only the leading large logarithms are calculated, and the higher order terms
are irrelevant.

At one loop, the physical masses and wave-function renormalizations are given by~\cite{Gasser:1984gg}
\begin{eqnarray}
m_\pi^2 &=& m_{\pi,0}^2\left[1+L(m_\pi)-\frac{1}{3}L(m_\eta)+\dots\right] \ ,  \\
m_K^2   &=& m_{K,0}^2\left[1+\frac{2}{3}L(m_\eta)+\dots\right] \ , \\
Z_\pi   &=& 1+\frac{4}{3}L(m_\pi)+\frac{2}{3}L(m_K)+\dots \ , \\
Z_K     &=& 1+\frac{1}{4}L(m_\pi)+\frac{1}{2}L(m_K)+\frac{1}{4}L(m_\eta)+\dots \ ,  \\
f_\pi &=& f\left[1-2L(m_\pi) - L(m_K) + \dots\right] \ , \\
f_K  &=& f\left[1-\frac{3}{4}L(m_\pi) - \frac{3}{2}L(m_K) - \frac{3}{4}L(m_\eta) + \dots\right]\ ,
\end{eqnarray}
for the pion and kaon fields, respectively. $L(m)$ is the chiral logarithm defined as:
\begin{equation}
L(m)\equiv\frac{m^2}{(4\pi f)^2}\ln\frac{m^2}{\mu_\chi^2} \ ,
\end{equation}
with $\mu_\chi$ the cutoff scale. The dots represent non-logarithm contributions from $\m{O}(p^4)$ and higher order Lagrangian terms. In this paper we focus only on the large chiral logarithmic corrections
and will not, for simplicity, include the dots explicitly in the results.

In the standard electroweak theory, there are 7 independent four-quarks operators which can be
classified into $(8,1)$, $(27, 1)$, and $(8,8)$. Define $\Theta = T^{ik}_{jl}\bar q_L^i\gamma_\mu
q_{Lj} \bar q^k_L \gamma^\mu q_{Ll}$, we can obtain four-independent quark operators with the
following tensor components
\begin{eqnarray}
\label{SMopeStart}
(27,1)_{3/2} &:& T^{31}_{21}=T^{31}_{12}=T^{13}_{21}=T^{13}_{12}=-T^{32}_{22}=-T^{23}_{22}=\frac{1}{2}\ ,\\
(27,1)_{1/2} &:& T^{31}_{21}=T^{31}_{12}=T^{13}_{21}=T^{13}_{12}=\frac{1}{2},T^{32}_{22}=T^{23}_{22}=1,
T^{33}_{23}=T^{33}_{32}=-\frac{3}{2}\ , \\
(8,1)_{1/2,S} &:&  T^{31}_{21} = T^{13}_{12} =  T^{31}_{12} = T^{13}_{21}  = \frac{1}{2}, T^{32}_{22}  = T^{23}_{22}= T^{33}_{23} = T^{33}_{32}= 1 \ , \nonumber \\
(8,1)_{1/2,A} &:&  T^{31}_{21} = T^{13}_{12} = - T^{31}_{12} = - T^{13}_{21}= \frac{1}{2} \ .
\end{eqnarray}
On the other hand, defining a (8,8) operator $\Theta = T^{ik}_{jl}\bar q_L^i\gamma_\mu
q_{Lj} \bar q^k_R \gamma^\mu q_{Rl}$, we have three quark operators with following tensor components,
\begin{eqnarray}
(8,8)_{3/2} &:& T^{31}_{21}=T^{31}_{12}=-T^{32}_{22}=1\ ,\\
(8,8)_{1/2,S}&:& T^{31}_{21}=T^{31}_{12}=T^{32}_{22}/2=-T^{33}_{23}/3=1 \ , \\
(8,8)_{1/2,A}&:& T^{31}_{21}=-T^{31}_{12}=-T^{33}_{23}=1 \ .
\end{eqnarray}
One can similarly defined other (8,8) operators with a different color indices contractions.
For the sake of convenience, we have broken the operators into representations of definite isospins.
This has the advantage of easily building up reducible operators from linear
combinations of these simple ones. For example, the SM electromagnetic penguin
operators $Q_{7,8}$:
\begin{equation}
Q_7=\frac{1}{2}\left[\Theta^{(8,8)}_{3/2}+\Theta^{(8,8)}_{1/2,A}\right] \ ,
\end{equation}
and $Q_8$ is similar to $Q_7$ but with different color indices contraction.

In ChPT, one can match the above QCD operators to the hadronic operators made of
Goldstone boson fields~\cite{Bernard:1985wf,Bijnens:1983ye,Cirigliano:1999pv}:
\begin{eqnarray}
\tilde \Theta^{(8,1)}_1 &\equiv& \tr\left[\Lambda\p_\mu\Sigma\p^\mu\Sigma^\dagger\right]\ , \nonumber \\
\tilde \Theta^{(8,1)}_2 &\equiv& \frac{8v}{f^2}\tr\left[\Lambda\Sigma M+\Lambda(\Sigma M)^\dagger \right] \ , \nonumber \\
\tilde \Theta^{(27,1)}_{\Delta I}  &\equiv& [T_{\Delta I}^{(27,1)}]_{kl}^{ij} (\Sigma\p_\mu\Sigma^\dagger)_{~i}^{k}(\Sigma\p^\mu\Sigma^\dagger)_{~j}^l \ , \nonumber \\
\tilde \Theta^{(8,8)}_{\Delta I}   &\equiv& [T_{\Delta I}^{(8,8)}]^{ij}_{kl} (\Sigma)^k_{~j}(\Sigma^\dagger)^l_{~i} \ , \nonumber\\
\tilde \Theta^{(\ovl{3},3)}        &\equiv& \tr[\Lambda\Sigma^\dagger] \ ,
\label{SMopeEnd}
\end{eqnarray}
where $\Lambda=\delta_{i,3}\delta_{j,2}$ and $T$'s are tensor structures defined above.
The expansions go like
\begin{eqnarray}
  \Theta^{(8,1)}_i  &=& \alpha_{1i}^{(8,1)}\tilde \Theta^{(8,1)}_1 +  \alpha_{2i}^{(8,1)}\tilde \Theta^{(8,1)}_2 + ... \nonumber \ ,\\ \Theta^{(27,1)}_{\Delta I}  &=& \alpha^{(27,1)}\tilde \Theta^{(27,1)}_{\Delta I} + ... \nonumber \ ,\\
  \Theta^{(8,8)}_{\Delta I} &=&  \alpha^{(8,8)}\tilde \Theta^{(8,8)}_{\Delta I} + ... \ ,
\end{eqnarray}
where $\alpha^{(L,R)}$s are ``Wilson coefficients" which are universal in different processes,
and dots represent higher dimensional operators. The subscript $i$ on the $(8,1)$ operator
indicates different quark operators in the same chiral representation,
including ones with two right-handed fields coupled to the singlet.

The one loop results of these operators in various processes can be found in~\cite{Bijnens:1984ec,Bijnens:1984qt,Golterman:1997st,Golterman:2000fw,Bijnens:1998mb,Cirigliano:1999pv}.  Due to the different definitions of the operators and the nonlinear meson fields,
there might be sign differences among these results.  Overall speaking, the (8,1) and (27,1)
operators dominate in the CP-conserve process. The (8,8) operators, corresponding to the $Q_{7,8}$ operators, play a significant role in $CP$-violation processes~\cite{Cirigliano:1999pv,Cirigliano:2001hs}.  The lowest-order mass-dependent
term $\Theta^{(8,1)}_2$ will vanish in physical process $K\rightarrow\pi\pi$ to all orders.
This property was pointed out by~\cite{Bernard:1985wf} first and has been well-studied by~\cite{Kambor:1989tz,Crewther:1985zt}.  We will come back to this issue later.

\subsection{Chiral Matching of New Operators}

Now we can proceed in constructing new hadronic operators for new interactions arising
from physics beyond SM. We label operators by the irreducible representatives
and their isospin quantum numbers.  Similar to the case in SM,
we define the effective operators at their lowest order as:
\begin{eqnarray}
\mbox {Dimension-6:\hspace{1.5cm}}& &\nonumber\\
\tilde{\Theta}^{(\ovl{6},6)}_{\Delta I} &=&  [T^{(\ovl{6},6)}_{\Delta I} ]_{kl}^{ij} (\Sigma^\dagger)_{~i}^k (\Sigma^\dagger)_{~j}^l \ , \\
\mbox {Dimension-7:\hspace{1.5cm}}& &\nonumber\\
\tilde{\Theta}^{(\ovl{15},3)}_{\Delta I} &=&  [T^{(\ovl{15},3)}_{\Delta I} ]_{kl}^{ij} (\Sigma\p^\mu\Sigma^\dagger)_{~i}^k (\p_\mu\Sigma^\dagger)_{~j}^l\ , \\
\tilde{\Theta}^{(6,3)}_{\Delta I} &=&
[T^{(6,3)}_{\Delta I} ]_{kl}^{ij} (\Sigma\p^\mu\Sigma^\dagger)_{~i}^k (\p_\mu\Sigma^\dagger)_{~j}^l\ , \\
\tilde{\Theta}^{(\ovl{3},15)}_{\Delta I} &=&  [T^{(\ovl{3},15)}_{\Delta I} ]_{kl}^{ij} (\p^\mu\Sigma^\dagger)_{~i}^k (\Sigma^\dagger\p_\mu\Sigma)_{~j}^l\ , \\
\tilde{\Theta}^{(\ovl{3},\ovl{6})}_{\Delta I} &=&  [T^{(\ovl{3},\ovl{6})}_{\Delta I} ]_{kl}^{ij} (\p^\mu\Sigma^\dagger)_{~i}^k (\Sigma^\dagger\p_\mu\Sigma)_{~j}^l \ .
\end{eqnarray}
We can also construct operators with one insertion of quark masses,
\begin{equation}
X_\pm^L \equiv (\Sigma M)\pm(\Sigma M)^\dagger,\;\;
X_\pm^R \equiv (M \Sigma)\pm(M \Sigma)^\dagger \ .
\end{equation}
They transform under $SU(3)_L\times SU(3)_R$ as:
\begin{equation}
X_\pm^L \rightarrow L X_\pm^L L^\dagger;\;\;X_\pm^R \rightarrow R X_\pm^R R^\dagger \ .
\end{equation}
With the insertion of $X_\pm$ we can build two additional sets of the dimension-7
operators at the lowest order:
\begin{equation}
\tilde{\Theta}'^{(L,R)}_{\Delta I,X_\pm} =  [T^{(L,R)}_{\Delta I} ]_{kl}^{ij} (X^L_\pm)_{~i}^k (\Sigma^\dagger)_{~j}^l \ ,
\end{equation}
for $(L,R)$ belong to $(\ovl{15},3)$ or $(6,3)$, and
\begin{equation}
\tilde{\Theta}'^{(L,R)}_{\Delta I,X_\pm} =  [T^{(L,R)}_{\Delta I} ]_{kl}^{ij} (\Sigma^\dagger)_{~i}^k (X^R_\pm)_{~j}^l \ ,
\end{equation}
for $(L,R)$ belong to $(\ovl{3},15)$ or $(\ovl{3},\ovl{6})$.  Therefore, the dimension-6 and -7 QCD operators should be matched to hadron operators as follows,
\begin{eqnarray}
\Theta_{\rm D6}^{(L,R)} & \rightarrow & \alpha^{(L,R)}\tilde{\Theta}^{(L,R)}\ , \\
\Theta_{\rm D7}^{(L,R)} & \rightarrow & \alpha^{(L,R)}\tilde{\Theta}^{(L,R)} + \alpha_{X_+}^{(L,R)}\tilde{\Theta}'^{(L,R)}_{X_+} + \alpha_{X_-}^{(L,R)}\tilde{\Theta}'^{(L,R)}_{X_-} \ ,
\end{eqnarray}
where higher-order terms have been omitted.

In SM we need operators with $X_+$ only since all QCD operators obey the CPS symmetry,
the CP transformation followed by an exchange of $s$ and $d$ quarks~\cite{Bernard:1985wf}.  However, the 4-quark operators derived from new physics do not necessarily have this symmetry, and hence we can have an
additional set of operators in the effective theory.

Just as the $(8,8)$ operators in SM, the $(\ovl{6},6)$ dimension-6 operators will contribute at $\mathcal{O}(p^0)$ order in
ChPT.  This set of operators can be derived from the Higgs (or some new heavy bosons)
exchange and will contribute to the CP violation phase in the same manner as $Q_7,Q_8$ in SM.
We will consider an example of applying our result later.

\begin{table}[!htb]
\caption{Tree level contributions from dimension-6 operators}
\begin{tabular*}{0.75\textwidth}{@{\extracolsep{\fill}} l*{3}{r}}
    \hline\hline
    & $K^0\rightarrow{\it Vacuum}$ & $K^+\rightarrow\pi^+$ & $K^0\rightarrow\pi^0\pi^0$\\
    \cline{2-2}\cline{3-3}\cline{4-4}
    $(L,R)_{\Delta I}$\hspace{1cm} & $b_0$ & $c_0$ & $d_0$\\
    \hline
    $(\ovl{6},6)_{3/2}$              & $0$ & $-4$ & $-8$ \\
    $(\ovl{6},6)_{1/2,S}$            & $-6$ & $-10$ & $16$ \\
    $(\ovl{6},6)_{1/2,A}$            & $-2$ & $-2$ & $0$  \\
    \hline\hline
    \label{D6Tree}
\end{tabular*}
\end{table}

When calculating CP conserving matrix elements, the new operators are usually negligible
 compared to the SM weak-interaction operators defined in Eqs.(\ref{SMopeStart})-(\ref{SMopeEnd}).  The new operators are mainly responsible for the CP-violating phase, and are worth investigating
 as the case in \cite{Cirigliano:1999pv,Cirigliano:2001hs}.
We observe that some of these new operators, notably $(\ovl{6},6),(\ovl{15},3),(6,3)$ and $(\ovl{3},15)$ have contributions in $\Delta I=3/2$ channel, in addition to $(27,1)$ and $(8,8)$ operators in SM. Furthermore, as we shall see, they all receive large chiral logarithmic corrections in one-loop ChPT. Therefore, the operators from new flavor theories beyond SM can help explain the $\Delta I=1/2$
selection rule and the direct CP violation parameter $\epsilon'$.

\subsection{Results at Tree Level}
In this subsection, we consider tree-level relations among the matrix elements of
the QCD operators in different states. These relations reflect chiral symmetry and can also
be derived using old-fashioned current algebra.

There are three processes that we are mainly interested in: $K^0\rightarrow{\it Vacuum}$, $K^+\rightarrow\pi^+$ and $K^0\rightarrow\pi^0\pi^0$. For $K^0 \rightarrow \pi^+\pi^-$, one can
obtained the matrix elements through angular momentum relation, as shown in Appendix.
At tree level, the dimension-7 momentum operators will not contribute to $K^0\rightarrow{\it Vacuum}$ process.  For the mass-dependent operators, the result will be either proportional to $(m_s-m_d)\sim m_{K,0}^2-m_{\pi,0}^2$ or $(m_s+m_d) \sim m_{K,0}^2$, by the lowest order expansion of $X_\pm$.
However, beyond tree level, the result will no longer be proportional to $(m_s-m_d)$ or $(m_s+m_d)$.

For dimension-6 operators in $({\bar 6}, 6)$, we have the tree level results:
\begin{eqnarray}
\langle0|\Theta_{D6}|K^0\rangle_{\rm Tree} & = & \frac{i\,b_0}{f}\alpha_{D6} \ , \\
\langle\pi^+|\Theta_{D6}|K^+\rangle_{\rm Tree} & = & \frac{c_0}{f^2}\alpha_{D6} \ , \\
\langle\pi^0\pi^0|\Theta_{D6}|K^0\rangle_{\rm Tree} & = & \frac{i\,d_0}{f^3}\alpha_{D6}\ ,
\end{eqnarray}
with $b_0$,$c_0$ and $d_0$ coefficients listed in Table \ref{D6Tree}, which are
different from different isospin projections. The corresponding
results for $\pi^+\pi^-$ final state can be obtained from relations in Appendix. The non-perturbative
coefficient $\alpha_{D6}$ is the same for different operators and final states.

\begin{table}
\caption{Tree level contributions from dimension-7 operators}
\begin{tabular*}{0.85\textwidth}{@{\extracolsep{\fill}} l*{8}{r}}\hline\hline
    & \multicolumn{2}{c}{$K^0\rightarrow{\it Vacuum}$} & \multicolumn{3}{c}{$K^+\rightarrow\pi^+$} & \multicolumn{3}{c}{$K^0\rightarrow\pi^0\pi^0$}\\
    \cline{2-3} \cline{4-6} \cline{7-9}
    \rule[-2mm]{0mm}{6mm}$(L,R)_{\Delta I}$ & $b'_{0,+}$ & $b'_{0,-}$ & $c'_0$ & $c'_{0,+}$ & $c'_{0,-}$ & $d'_0$ & $d'_{0,+}$ & $d'_{0,-}$\\
    \hline
    $(\ovl{15},3)_{3/2}$        & $0$ & $0$ & $-8 $ & $0$ & $2$ & $8 $ & $-2$ & $-2$\\
    $(\ovl{15},3)_{1/2,S}$      & $9/2$ & $3/2$ & $-8 $ & $-3/2$ & $2$ & $8 $ & $-2$ & $-2$\\
    $(\ovl{15},3)_{1/2,A}$      & $-1/2$ & $1/2$ & $0 $ & $-1/2$ & $0$ & $8 $ & $-2$ & $-2$\\
    $(\ovl{15},3)_{1/2}$        & $1/2$ & $1/2$ & $0 $ & $-1/2$ & $0$  & $0 $ & $0$ & $0$\\
    $(6,3)_{3/2}$               & $0$ & $0$ & $0 $ & $0$ & $0$ & $8 $ & $-2$ & $-2$\\
    $(6,3)_{1/2}$               & $3/2$ & $3/2$ & $12 $ & $-3/2$ & $-3$ & $16 $ & $-4$ & $-4$\\
    $(\ovl{3},15)_{3/2,L}$      & $0$ & $0$ & $4 $  & $0$ & $-1$ & $0 $ & $0$ & $0$\\
    $(\ovl{3},15)_{3/2,R}$      & $0$ & $0$ & $4 $  & $0$ & $-1$ & $8 $ & $2$ & $-2$\\
    $(\ovl{3},15)_{1/2,L}$      & $-3/2$ & $-3/2$ & $4 $ & $3/2$ & $-1$ & $0 $ & $0$ & $0$\\
    $(\ovl{3},15)_{1/2,R}$      & $-3/2$ & $3/2$ & $4 $ & $-3/2$ & $-1$ & $-16 $ & $2$ & $-2$\\
    $(\ovl{3},\ovl{6})_{1/2,L}$ & $3/2$ & $-1/2$ & $4 $ & $1/2$ & $-1$ & $0 $ & $0$ & $0$\\
    $(\ovl{3},\ovl{6})_{1/2,R}$ & $-1/2$ & $1/2$ & $-4 $ & $-1/2$ & $1$ & $0 $ & $2$ & $-2$\\
    \hline\hline
\end{tabular*}
\label{D7Tree}
\end{table}

Similarly, the tree-level matrix elements for dimension-7 operators are:
\begin{eqnarray}
\langle0|\Theta_{D7}|K^0\rangle_{\rm Tree} &
= & \frac{4iv}{f^3} \left[b'_{0,+}(m_s-m_d)\alpha_{D7,+} \,+b'_{0,-}(m_s+m_d)\alpha_{D7,-}\right]\nonumber\\
 & = &  \frac{i}{f} \left[b'_{0,+}(m_{K,0}^2-m_{\pi,0}^2)\alpha_{D7,+} \,+b'_{0,-}m_{K,0}^2\alpha_{D7,-}\right] \ , \\
\langle\pi^+|\Theta_{D7}|K^+\rangle_{\rm Tree} & = & \frac{m_{M,0}^2}{f^2}[c'_0\alpha_{D7} + c'_{0,+}\alpha_{D7,+}
   + c'_{0,-}\alpha_{D7,-} ]  \ , \\
\langle\pi^0\pi^0|\Theta_{D7}|K^0\rangle_{\rm Tree} & = & \frac{ m_K^2}{f^3}[d'_0\alpha_{D7} + d'_{0,+}\alpha_{D7,+}   +  d'_{0,-}\alpha_{D7,-}]  \ ,
\end{eqnarray}
with coefficients listed in Table \ref{D7Tree}. Here the non-perturbative coefficients $\alpha's$
are different for different chiral representations. From the above equations, it is clear that one can obtained the two-pion
matrix elements from the vacuum and one-pion ones, if only one of the $X_+$ and  $X_-$ types of operators is present,
such as in the SM case.

\section{Chiral Logarithms at One-Loop}

ChPT calculations of the kaon-decay matrix elements up to higher chiral orders are needed
for understanding the size of chiral corrections and for extrapolating matrix elements
from unphysical quark masses to physical ones. In lattice calculations, unphysically large
quark masses are usually used to make calculations feasible. Then one needs
to extrapolate the matrix elements to the physical region. In this section,
we calculate the large chiral logarithms of the dimension-6 and 7 operators
for the process $K\rightarrow0$, $K\rightarrow\pi$ and $K\rightarrow\pi\pi$ hoping
to get the leading corrections as the function of quark masses.

In our calculations, we have made the simplifying assumption $m_u=m_d$.
For $\langle0|\mathcal{O}|K^0\rangle$ matrix element, we have kept
all the Goldstone boson masses independent.
For the matrix element $\langle\pi^+|\mathcal{O}|K^+\rangle$, we utilize a common mass $m_M$ for all the mesons to conserve momentum.  In the calculation of $\langle\pi^0\pi^0|\mathcal{O}|K^0\rangle$ matrix elements the pion masses are neglected. Since $m_\pi^2/m_K^2\approx10^{-1}$ with physical pion and kaon masses, this is a reasonable approximation for the physical processes.

\subsection{$K^0\rightarrow{\it Vacuum}$}
The diagram we need to consider is shown in Fig. \ref{K2vacuum} below, where and henceforth
the square dot represents an effective weak interaction operator while the round dots represent
strong interaction insertions.  We have not shown the wave function renormalization diagrams,
but they have to be included in the final result.

\begin{figure}[hbt]
\begin{center}
\includegraphics[scale=.7]{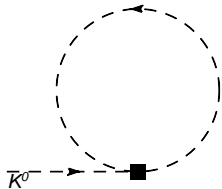}
\end{center}
\caption{Feynman diagram for $K^0\rightarrow{\it Vacuum}$ at one loop.}
\label{K2vacuum}
\end{figure}

For dimension-6 operators, the results for $\langle0|\mathcal{O}|K^0\rangle$ up to one-loop
can be written as:
\begin{equation}
\langle0|\Theta_{D6}|K^0\rangle=\frac{i\alpha_{D6}}{f}\left[b_0+b_\eta L(m_\eta^2)+b_K L(m_K^2)+b_\pi L(m_\pi^2)\right] \ ,
\end{equation}
with coefficients listed in Table \ref{D6}. Note that we have three different chiral logarithms
corresponding to eta, kaon, and pion, respectively. The isospin-3/2 operator does
contribute for the obvious reason.

\begin{table}[!htb]
\caption{One loop contributions from dimension-6 operators}
\begin{tabular*}{0.75\textwidth}{@{\extracolsep{\fill}} l*{6}{r}}
    \hline\hline
     & \multicolumn{3}{c}{$K^0\rightarrow{\it Vacuum}$} & $K^+\rightarrow\pi^+$ & $K^0\rightarrow\pi^0\pi^0$ \\
    \cline {2-4} \cline{5-5} \cline{6-6}
    $(L,R)_{\Delta I}$ & $b_\eta$ & $b_K$ & $b_\pi$ & $c_M$ & $d_K$\\  \hline
    $(\ovl{6},6)_{3/2}$   & $0$ & $0$ & $0$ & $112/3$ & $80/9$\\
    $(\ovl{6},6)_{1/2,S}$ & $1/2$ & $33$ & $57/2$ & $196/3$ & $-1024/9$\\
    $(\ovl{6},6)_{1/2,A}$ & $25/6$ & $15$ & $3/2$ & $28/3$ & $-4$\\
    \hline\hline
\end{tabular*}
\label{D6}
\end{table}

For dimension-7 operators, the results are more complicated,
\begin{eqnarray}
\langle0|\Theta_{D7}|K^0\rangle & = & \langle0|\Theta_{D7}|K^0\rangle_{\rm Tree}+\frac{i}{f}\Big\{\left[b'_\eta m_\eta^2 L(m_\eta^2)+b'_K m_K^2L(m_K^2)+b'_\pi m_\pi^2L(m_\pi^2)\right]\!\cdot\alpha_{D7}\nonumber\\
 & & +\sum_{\pm}\left[(b_{\eta,K}' m_K^2+b_{\eta,\pi}' m_\pi^2)L(m_\eta^2)+(b_{K,K}' m_K^2+b_{K,\pi}' m_\pi^2) L(m_K^2)\right.\nonumber\\
 & &\left.+(b_{\pi,K} m_K^2+b_{\pi,\pi}' m_\pi^2)L(m_\pi^2)\right]\!\cdot\alpha_{D7,\pm}\Big\} \ ,
\end{eqnarray}
where each chiral logarithms now have different meson mass factors.
The coefficients are listed in Tables \ref{D7Mom} and \ref{D7MassK20}.

\subsection{$K^+\rightarrow\pi^+$}

\begin{figure}[hbt]
\begin{center}
\includegraphics[scale=.7]{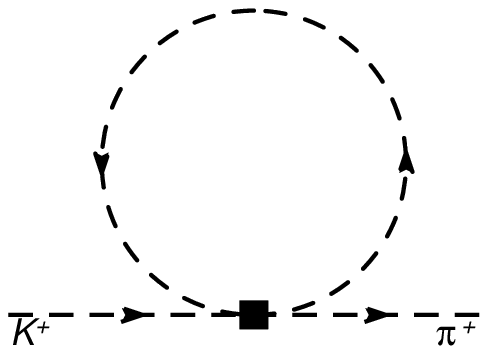}
\hspace{2cm}
\includegraphics[scale=.7]{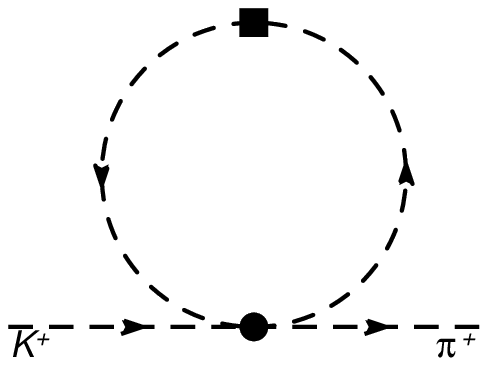}
\end{center}
\caption{Feynman diagrams for $K^+\rightarrow\pi^+$ at one-loop.}
\label{K2PFig}
\end{figure}

For the $K^+\rightarrow\pi^+$ matrix elements, we utilize a common mass for mesons $m_M^2=m_\pi^2=m_K^2$ in the calculation.
The Feynman diagrams are shown in Fig. \ref{K2PFig}. The matrix elements up to the leading
chiral logarithms are:
\begin{eqnarray}
\langle \pi^+|\Theta_{D6}|K^+\rangle    &  = &\frac{\alpha_{D6}}{f^2} \left[c_0 + c_M L(m_M^2)\right] \ ,\\
\langle \pi^+|\Theta_{D7}|K^+\rangle    &  = & \langle \pi^+|\Theta_{D7}|K^+\rangle_{\rm Tree}\nonumber\\
 & & +\frac{m_{M,0}^2}{f^2}L(m_M^2) \Big[c'_M \alpha_{D7} + c'_{M,+} \alpha_{D7,+}+ c'_{M,-} \alpha_{D7,-}\Big] \ ,
\end{eqnarray}
for dimension-6 and 7 operators, respectively. The coefficients are listed in Tables \ref{D6} and \ref{D7Mass}.

\subsection{$K^0\rightarrow\pi^0\pi^0$}

\begin{figure}[!hbt]
\begin{center}
\includegraphics[scale=.7]{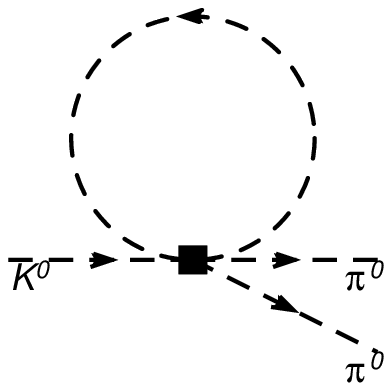}
\hspace{5mm}
\includegraphics[scale=.7]{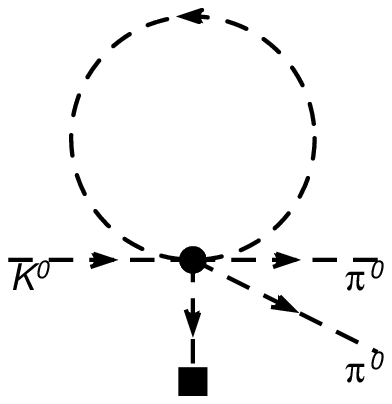}
\hspace{5mm}
\includegraphics[scale=.7]{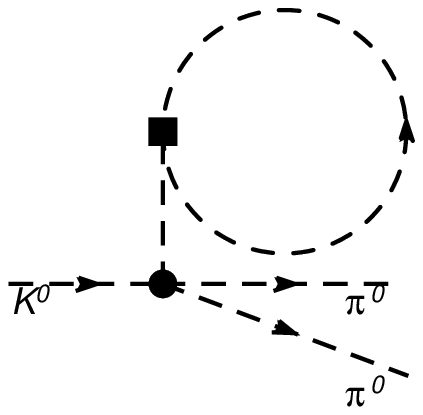}
\hspace{5mm}
\includegraphics[scale=.7]{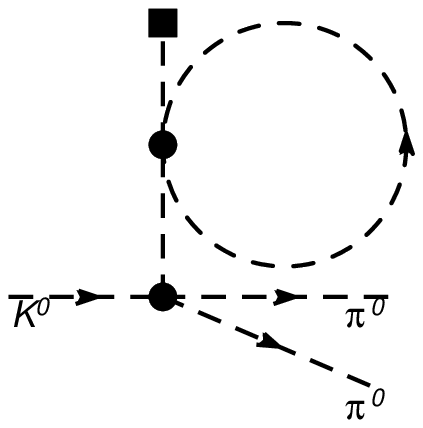}\\
\mbox{\vspace*{5mm}}\\
\includegraphics[scale=.7]{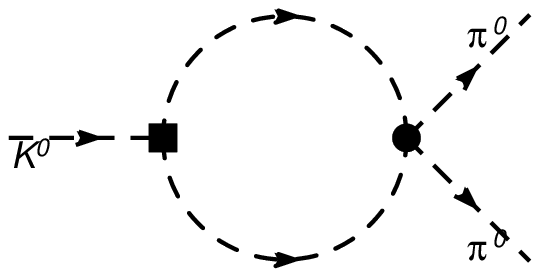}
\hspace{5mm}
\includegraphics[scale=.7]{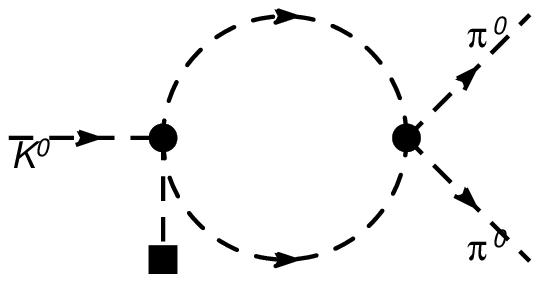}
\hspace{5mm}
\includegraphics[scale=.7]{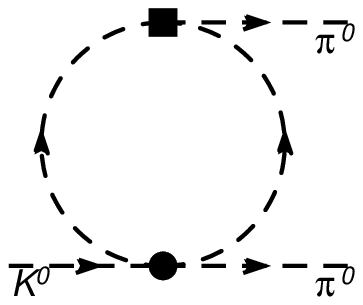}
\hspace{5mm}
\includegraphics[scale=.7]{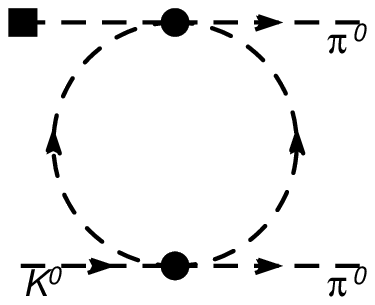}
\caption{Feynman diagrams for $K^0\rightarrow\pi^0\pi^0$ at one-loop ChPT.}
\label{K2PPFig}
\end{center}
\end{figure}

For $K^0\rightarrow\pi^0\pi^0$, we neglect the pion mass in the calculation. The diagrams we need to consider for
are shown in Fig. \ref{K2PPFig}. The results are:
\begin{eqnarray}
\langle \pi^0\pi^0|\Theta_{D6}|K^0\rangle    & = & \frac{i\alpha_{D6}}{f^3} \left[d_0 + d_K L(m_K^2)\right] \ ,\\
\langle \pi^0\pi^0|\Theta_{D7}|K^0\rangle    & = & \langle\pi^0\pi^0|\Theta_{D7}|K^0\rangle_{\rm Tree}\nonumber\\
 & & +\frac{m_K^2}{f^2}L(m_K^2) \Big[d'_M \alpha_{D7} + d'_{M,+} \alpha_{D7,+}+ d'_{M,-} \alpha_{D7,-}\Big] \ ,
\end{eqnarray}
with coefficients listed in Tables \ref{D6}, \ref{D7Mass}.

\begin{table}[!htb]
\caption{One loop contributions from dimension-7 operators (I)}
\begin{tabular*}{0.75\textwidth}{@{\extracolsep{\fill}} l*{3}{r}}
    \hline\hline
     & \multicolumn{3}{c}{$K^0\rightarrow{\it Vacuum}$} \\
    \cline {2-4}
    $(L,R)_{\Delta I}$ & $b'_\eta$ & $b'_K$ & $b'_\pi$ \\
    \hline
    $(\ovl{15},3)_{1/2,S}$      & $0$ & $-12$ & $12$ \\
    $(\ovl{15},3)_{1/2,A}$      & $4$ & $-4$ & $0$  \\
    $(\ovl{15},3)_{1/2}$        & $-4$ & $4$ & $0$  \\
    $(6,3)_{1/2}$               & $6$ & $12$ & $-18$ \\
    $(\ovl{3},15)_{1/2,L}$      & $18$ & $-12$ & $-6$ \\
    $(\ovl{3},15)_{1/2,R}$      & $6$ & $-12$ & $6$ \\
    $(\ovl{3},\ovl{6})_{1/2,L}$ & $2$ & $4$ & $-6$ \\
    $(\ovl{3},\ovl{6})_{1/2,R}$ & $2$ & $4$ & $-6$ \\
    \hline\hline
    \label{D7Mom}
\end{tabular*}
\end{table}

\begin{table}[!htb]
\caption{One loop contributions from dimension-7 operators (II)}
\begin{tabular*}{0.85\textwidth}{@{\extracolsep{\fill}} l*{6}{r}}
    \hline\hline
    & \multicolumn{6}{c}{$K^0\rightarrow{\it Vacuum}$}\\
    \cline{2-7}
    \rule[-3mm]{0mm}{8mm}$(L,R)_{\Delta I,X\pm}$ & $b'_{\eta,K}$ & $b'_{\eta,\pi}$ & $b'_{K,K}$ & $b'_{K,\pi}$ & $b'_{\pi,K}$ & $b'_{\pi,\pi}$\\
    \hline
    $(\ovl{15},3)_{1/2,S,X_+}$ & $-43/8$ & $27/8$ & $-123/4$ & $111/4$ & $-51/8$ & $99/8$ \\
    $(\ovl{15},3)_{1/2,S,X_-}$ & $7/8$ & 0 & $-33/4$ & 0 & $-33/8$ & $-3$ \\
    $(\ovl{15},3)_{1/2,A,X_+}$ & $25/24$ & $-17/24$ & $27/4$ & $-15/4$ & $-21/8$ & $-3/8$ \\
    $(\ovl{15},3)_{1/2,A,X_-}$ & $-25/24$ & $1/3$ & $-3/4$ & $0$ & $-27/8$ & $0$ \\
    $(\ovl{15},3)_{1/2,X_+}$ & $-3/8$ & $17/24$ & $-15/4$ & $15/4$ & $-3/8$ & $3/8$ \\
    $(\ovl{15},3)_{1/2,X_-}$ & $-3/8$ & $-1/3$ & $-15/4$ & $0$ & $-3/8$ & $0$ \\
    $(6,3)_{1/2,X_+}$ & $3/2$ & $5/8$ & $-21/4$ & $9/4$ & $-57/8$ & $45/8$ \\
    $(6,3)_{1/2,X_-}$ & $3/2$ & $1/2$ & $-21/4$ & $0$ & $-57/8$ & $9/2$ \\
    $(\ovl{3},15)_{1/2,L,X_+}$   & $25/8$ & $-5/8$ & $45/4$ & $-33/4$ & $9/8$ & $-45/8$ \\
    $(\ovl{3},15)_{1/2,L,X_-}$   & $25/8$ & $1/2$ & $45/4$ & $0$ & $9/8$ & $-3/2$ \\
    $(\ovl{3},15)_{1/2,R,X_+}$   & $9/8$ & $-5/8$ & $33/4$ & $-33/4$ & $33/8$ & $-45/8$ \\
    $(\ovl{3},15)_{1/2,R,X_-}$   & $-9/8$ & $1/2$ & $-33/4$ & $0$ & $-33/8$ & $-3/2$ \\
    $(\ovl{3},\ovl{6})_{1/2,L,X_+}$  & $-43/24$ & $13/8$ & $-45/4$ & $33/4$ & $-9/8$ & $21/8$ \\
    $(\ovl{3},\ovl{6})_{1/2,L,X_-}$  & $-7/24$ & $1/2$ & $15/4$ & $0$ & $3/8$ & $-3/2$ \\
    $(\ovl{3},\ovl{6})_{1/2,R,X_+}$  & $3/8$ & $-5/24$ & $3/4$ & $-3/4$ & $27/8$ & $-15/8$ \\
    $(\ovl{3},\ovl{6})_{1/2,R,X_-}$  & $-3/8$ & $1/6$ & $-3/4$ & $0$ & $-27/8$ & $3/2$ \\
    \hline\hline
    \label{D7MassK20}
\end{tabular*}
\end{table}

\begin{table}
\caption{One loop contributions from dimension-7 operators (III)}
\begin{tabular*}{0.85\textwidth}{@{\extracolsep{\fill}} l*{6}{r}}\hline\hline
     & \multicolumn{3}{c}{$K^+\rightarrow\pi^+$} & \multicolumn{3}{c}{$K^0\rightarrow\pi^0\pi^0$} \\
    \cline{2-4} \cline{5-7}
    \rule[-2mm]{0mm}{6mm}$(L,R)_{\Delta I}$ & $(c'_M)$ & $(c'_M)_{X_+}$ & $(c'_M)_{X_-}$ & $(d'_K)$ & $(d'_K)_{X_+}$ & $(d'_K)_{X_-}$\\
    \hline
    $(\ovl{15},3)_{3/2}$     & $208/3$ & $3/4$ & $-52/3$ & $8$ & $-2/9$ & $-2/9$\\
    $(\ovl{15},3)_{1/2,S}$   & $196/3$ & $28/3$ & $-49/3$ & $-640/9$ & $214/9$ & $160/9$\\
    $(\ovl{15},3)_{1/2,A}$   & $-4/3$ & $8/3$ & $1/3$ & $-352/9$ & $70/9$ & $88/9$\\
    $(\ovl{15},3)_{1/2}$     & $-4/3$ & $8/3$ & $1/3$ & $-8$ & $2$ & $2$\\
    $(6,3)_{3/2}$            & $0$ & $0$ & $0$ & & $8/9$ $-2/9$ & $-2/9$\\
    $(6,3)_{1/2}$            & $-80$ & $12$ & $9$ & $-848/9$ & $212/9$ & $212/9$\\
    $(\ovl{3},15)_{3/2,L}$   & $-16$ & $2/3$ & $26/3$ & $0$ & $0$ & $0$\\
    $(\ovl{3},15)_{3/2,R}$   & $-16$ & $2/3$ & $26/3$ & $-16$ & $2/9$ & $-2/9$\\
    $(\ovl{3},15)_{1/2,L}$   & $0$ & $-22/3$ & $29/3$ & $0$ & $-3$ & $-3$\\
    $(\ovl{3},15)_{1/2,R}$   & $-32$ & $26/3$ & $23/3$ & $128$ & $-133/9$ & $133/9$\\
    $(\ovl{3},\ovl{6})_{1/2,L}$  & $-80/3$ & $-4$ & $3$ & $-8$ & $5$ & $-3$\\
    $(\ovl{3},\ovl{6})_{1/2,R}$  & $80/3$ & $4$ & $-3$ & $0$ & $-97/9$ & $97/9$\\
    \hline\hline
\end{tabular*}
\label{D7Mass}
\end{table}

There are several comments we would like to make.  First, the quark-mass-dependent operators (operators constructed with $X_\pm$)
contribute in the physical process $K\rightarrow\pi\pi$, in contrary to $\Theta_2^{(8,1)}$ in the SM case.
The reason is similar to that for the higher-order operators of (27,1)~\cite{Bijnens:1998mb}: they cannot be expressed as a
total divergence by the equations of motion.  Similarly, these new operators will not act as a generator for
rotation in $s-d$ plane like the $\Theta_2^{(8,1)}$ does~\cite{Bernard:1985wf,Kambor:1989tz,Crewther:1985zt}.
Therefore the one-loop matrix elements of $K\rightarrow0$ will no longer be proportional to $(m_s \pm m_d)$ as
they did at tree level.

Second, the masses appearing in our result are either bare masses or the renormalized one depending on the processes.
For the unphysical processes $K\to0$ and $K\to\pi$, we use the bare masses, whereas for the physical $K\to \pi\pi$,
the one-loop renormalized mass is implied. It makes the comparison to the experimental result feasible.

Third, in~\cite{Buchler:2005xn} the author claimed that infrared-sensitive terms like $m_K^2\log m_\pi^2$,
which diverges in the $m_\pi\rightarrow0$ limit, will emerge in the $K\rightarrow\pi\pi$ matrix element.
We have checked the result by keeping pion masses explicit in our calculations, and found that all such terms canceled
when summing all the diagrams. Therefore it is safe to take the limit $m_\pi\rightarrow0$.

Finally, there are a large number of unknown non-perturbative coefficients in the new operators.
For dimension-6 operators, the traditional way of determining these coefficients
by calculating simple processes like $K^+\rightarrow\pi^+$ is suffice.
In dimension-7 cases, however, the two simple processes $K^0\rightarrow0$ and $K^+\rightarrow\pi^+$ are not enough
to determine all the coefficients, unless there is the so-called CPS symmetry.
Adding other simple processes like $K^0\rightarrow\pi^0$ and
$K^0\rightarrow\eta$ will not improve the situation since they are not independent in the $SU(3)$
limit we are working on.  We can in principle get more relationships when away from $SU(3)$ chiral
and isospin symmetries, but many more new coefficients will enter as well, and then
we need even more relationships to determine all the coefficients.  Therefore we could either rely on
some model-dependent assumptions or calculate more complicate processes on lattice directly.
In any case, the ChPT calculations can serve as a check
for relations among coefficients from lattice or other non-perturbative model calculations.

\section{Conclusion}

The standard model calculations for the direct CP violation in non-leptonic kaon decay
have not been entirely settled due to large cancelations between different matrix elements.
Therefore, there is a considerable interest in understanding this phenomena from
beyond standard model physics. However, we do not know yet what form the
new physics will take, either supersymmetry, left-right symmetry, large extra dimensions,
or little Higgs, or something else. Presumably, the Large Hadron Collider will help
us to identify it in the next few years.

In this paper, we aim to study a general effective theory for non-leptonic kaon decay which has
its origin from beyond SM physics. We systematically classify the dimension-5, 6 and 7
quark and gluon operators according to their chiral structures.
Using chiral symmetry, we derive tree-level relations between
the matrix elements involving zero, one and two pions. This is useful
because lattice calculations of multiparticle matrix elements
are much harder than these for few particles. We have also calculated the leading chiral
logarithmic behavior of these operators in ChPT. The result again will be useful for calculating matrix
elements of these operators on lattice. We have not consider them in quenched
QCD formulations, as the rapid progress in lattice QCD calculations makes queched studies
much less useful than the past.

We thank J. Bijnens and M. Golterman for useful correspondences.
This work was partially supported by the U. S. Department of Energy via grant DE-FG02-93ER-40762. H. W. Ke acknowledges a scholarship support from China's Ministry of Education.
\newpage

\appendix
\section{Leading Chiral-Logarithms in SM Operators}

The leading chiral-logarithms in SM operators have been calculated by many authors \cite{Bijnens:1984qt,Bijnens:1984ec,Bijnens:1998mb,Cirigliano:1999pv,Cirigliano:2001hs}, and for completeness we list the result here.  Notice that the results quoted in Eqs.~(80) and (81) in \cite{Blum:2001xb}
contain sign errors.  The result for $K^0\rightarrow\pi^0\pi^0$ here is different from
that in \cite{Bijnens:1984qt}, as pointed out in \cite{Buchler:2005xn}.  Results for $K^0\rightarrow{\it Vacuum}$ and $K^+\rightarrow\pi^+$ are
presented in terms of bare masses and couplings, while for $K^0\rightarrow\pi^0\pi^0$ we use the physical mass.

The operators we use are defined in the main
body of the paper (Eqs. \ref{SMopeStart} - \ref{SMopeEnd}). We first consider the matrix elements between $K^0$ and the vacuum,
\begin{eqnarray}
\langle 0|\Theta^{(8,1)}|K^0\rangle & = & \frac{2i\alpha^{(8,1)}_1}{f}\left[m_\eta^2 L(m_\eta)+2m_K^2 L(m_K) -3m_\pi^2 L(m_\pi)\right]\nonumber\\
   & &+\frac{4i\alpha^{(8,1)}_2}{f}(m_{K,0}^2-m_{\pi,0}^2) \left[1-\frac{1}{12}L(m_\eta) -\frac{3}{2}L(m_K) -\frac{3}{4}L(m_\pi)\right] \ ,\\
\langle 0|\Theta^{(27,1)}_{1/2}|K^0\rangle  & = & \frac{6i\alpha^{(27,1)}}{f}\left[3m_\eta^2 L(m_\eta)-4m_K^2 L(m_K)+m_\pi^2 L(m_\pi)\right] \ ,\\
\langle 0|\Theta^{(8,8)}_{1/2,A}|K^0\rangle & = & \frac{12i\alpha^{(8,8)}}{f}\left[ L(m_K) - L(m_\pi)\right] \ ,\\
\langle 0|\Theta^{(8,8)}_{1/2,S}|K^0\rangle & = & -\frac{12i\alpha^{(8,8)}}{f}\left[1 - \frac{3}{4}L(m_\eta) - \frac{13}{2}L(m_K) - \frac{7}{4}L(m_\pi)\right] \ , \\
\langle 0|\Theta^{(\ovl{3},3)}|K^0\rangle & = & -\frac{2i\alpha^{(\ovl{3},3)}}{f}\left[1-\frac{1}{12}L(m_\eta) -\frac{3}{2}L(m_K) -\frac{3}{4}L(m_\pi)\right]
\end{eqnarray}
where $f$ is the bare meson decay constant, and $m_{\pi,0}$,$m_{K,0}$ are bare masses of mesons. Due to the isospin conservation, only $I=1/2$ part of the operator can contribute.

For $K^+\rightarrow\pi^+$ matrix elements, we apply a common mass $m_M$ for all the mesons. Therefore the momentum is conserved in the process.
\begin{eqnarray}
\langle \pi^+|\Theta^{(8,1)}|K^+\rangle & = & \frac{4m_{M,0}^2}{f^2}\left\{\alpha^{(8,1)}_1\left[1+\frac{1}{3}L(m_M)\right]-\alpha^{(8,1)}_2\left[1+2L(m_M)\right]\right\} \ ,\\
\langle \pi^+|\Theta^{(27,1)}_{3/2}|K^+\rangle & = & \langle \pi^+|\Theta^{(27,1)}_{1/2}|K^+\rangle \nonumber\\
 & = & -\frac{4m_{M,0}^2\alpha^{(27,1)}}{f^2}\left[1-\frac{34}{3}L(m_M)\right] \ ,\\
\langle \pi^+|\Theta^{(8,8)}_{3/2}|K^+\rangle & = & \frac{4\alpha^{(8,8)}}{f^2}\left[1 -8L(m_M)\right]  \ ,\\
\langle \pi^+|\Theta^{(8,8)}_{1/2,A}|K^+\rangle & = & \frac{8\alpha^{(8,8)}}{f^2}\left[1 - 5L(m_M)\right]  \ ,\\
\langle \pi^+|\Theta^{(8,8)}_{1/2,S}|K^+\rangle & = & \frac{4\alpha^{(8,8)}}{f^2}\left[1 - 8L(m_M)\right] \ , \\
\langle \pi^+|\Theta^{(\ovl{3},3)}|K^+\rangle & = & \frac{2\alpha^{(\ovl{3},3)}}{f^2}\left[1+2L(m_M)\right] \ .
\end{eqnarray}
This result is useful in lattice calculations where the pion mass can be adjusted through
quark mass parameters. The $K^0 \rightarrow \pi^0$ matrix elements can be obtained from the above by using,
\begin{eqnarray}
\langle \pi^0|\mathcal{O}_{\Delta I=1/2}|K^0\rangle & = & -\sqrt{\frac{1}{2}} \langle \pi^+|\mathcal{O}_{\Delta I=1/2}|K^+\rangle \ ,\nonumber\\
\langle \pi^0|\mathcal{O}_{\Delta I=3/2}|K^0\rangle & = & \sqrt{2}\langle \pi^+|\mathcal{O}_{\Delta I=3/2}|K^+\rangle \ .
\label{Kpi}
\end{eqnarray}

Finally, for $K\rightarrow\pi\pi$, we take the limit $m_\pi \rightarrow 0$ and keep the kaon mass dependency only,
\begin{eqnarray}
\langle \pi^0\pi^0|\Theta^{(8,1)}|K^0\rangle   & = & \frac{4i\alpha^{(8,1)}_1m_K^2}{f^3}\left[1-\frac{5}{4}L(m_K)\right] \ , \\
\langle \pi^0\pi^0|\Theta^{(27,1)}_{3/2}|K^0\rangle  & = & \frac{8i\alpha^{(27,1)}m_K^2}{f^3}\left[1-\frac{3}{2}L(m_K)\right] \ , \\
\langle \pi^0\pi^0|\Theta^{(27,1)}_{1/2}|K^0\rangle  & = & -\frac{4i\alpha^{(27,1)}m_K^2}{f^3}\left[1-15L(m_K)\right]\ , \\
\langle \pi^0\pi^0|\Theta^{(8,8)}_{3/2}|K^0\rangle & = & \frac{8i\alpha^{(8,8)}}{f^3}\left[1 + L(m_K)\right] \ , \\
\langle \pi^0\pi^0|\Theta^{(8,8)}_{1/2,A}|K^0\rangle & = & -\frac{8i\alpha^{(8,8)}}{f^3}\left[1 - \frac{7}{2}L(m_K)\right] \ ,  \\
\langle \pi^0\pi^0|\Theta^{(8,8)}_{1/2,S}|K^0\rangle & = & \frac{8i\alpha^{(8,8)}}{f^3}\left[1 - \frac{19}{2}L(m_K)\right] \ .
\end{eqnarray}
Here the physical mass of kaon is used.  Note that the weak mass operator, $\Theta^{(8,1)}_2$, will not contribute
to the $K\rightarrow\pi\pi$ matrix element as being pointed out in \cite{Bijnens:1983ye,Bernard:1985wf,Kambor:1989tz,Crewther:1985zt}.

The matrix elements for the final state $|\pi^+\pi^-\rangle$ is related to the above ones simply by
\begin{eqnarray}
   A_{+-} &=& \frac{1}{\sqrt{3}}(A_2 + \sqrt{2} A_0) \ ,\nonumber \\
   A_{00} &=& \sqrt{\frac{2}{3}}(-\sqrt{2} A_2 + A_0) \ .
\end{eqnarray}
Compared with the angular momentum relation, the $A_0$ amplitude has a factor of $-\sqrt{2}$. The minus sign arises from
the definition of $\pi^+ = (\pi^1 + i\pi^2)/\sqrt{2}$ which has a different sign from the usual spherical tensor definition.
The $\sqrt{2}$ accounts for the identical particle nature of two $\pi^0$'s, which is usually accounted from by a factor of 1/2
in the final state phase space. From the above relation, we derive:
\begin{eqnarray}
\langle \pi^+\pi^-|\mathcal{O}_{\Delta I=1/2}|K^0\rangle & = & \langle \pi^0\pi^0|\mathcal{O}_{\Delta I=1/2}|K^0\rangle \ ,\nonumber\\
\langle \pi^+\pi^-|\mathcal{O}_{\Delta I=3/2}|K^0\rangle & = & -\frac{1}{2}\langle \pi^0\pi^0|\mathcal{O}_{\Delta I=3/2}|K^0\rangle \ .
\label{Kpipi}
\end{eqnarray}
Using the relation (\ref{Kpi}) and (\ref{Kpipi}), it is easy to check that the result for (8,8) operators is consistent with
that in Ref. \cite{Cirigliano:1999pv}.


\begin{thebibliography}{99}
\bibitem{Christenson:1964fg}
  J.~H.~Christenson, J.~W.~Cronin, V.~L.~Fitch and R.~Turlay,
  Phys.\ Rev.\ Lett.\  {\bf 13}, 138 (1964).

\bibitem{CPV:Exp}
  A.~Alavi-Harati {\it et al.}  [KTeV Collaboration],
  Phys.\ Rev.\  D {\bf 67}, 012005 (2003)
  [Erratum-ibid.\  D {\bf 70}, 079904 (2004)]
  [arXiv:hep-ex/0208007];
  J.~R.~Batley {\it et al.}  [NA48 Collaboration],
  Phys.\ Lett.\  B {\bf 544}, 97 (2002)
  [arXiv:hep-ex/0208009];
  G.~D.~Barr {\it et al.}  [NA31 Collaboration],
  Phys.\ Lett.\  B {\bf 317}, 233 (1993);
  L.~K.~Gibbons {\it et al.},
  Phys.\ Rev.\ Lett.\  {\bf 70}, 1203 (1993).

\bibitem{PDG}
W.-M.Yao et al. (Particle Data Group), J. Phys. G 33, 1 (2006)
and 2007 partial update for the 2008 edition


\bibitem{Wolfenstein:1964ks}
  L.~Wolfenstein,
  Phys.\ Rev.\ Lett.\  {\bf 13}, 562 (1964).

\bibitem{Kobayashi:1973fv}
  M.~Kobayashi and T.~Maskawa,
  Prog.\ Theor.\ Phys.\  {\bf 49}, 652 (1973).


\bibitem{Buras:2003zz}
  A.~J.~Buras and M.~Jamin,
  JHEP {\bf 0401}, 048 (2004)
  [arXiv:hep-ph/0306217].



\bibitem{Ciuchini:2000zz}
  M.~Ciuchini and G.~Martinelli,
  Nucl.\ Phys.\ Proc.\ Suppl.\  {\bf 99B}, 27 (2001)
  [arXiv:hep-ph/0006056].

\bibitem{Ciuchini:1997kd}
  M.~Ciuchini,
  Nucl.\ Phys.\ Proc.\ Suppl.\  {\bf 59}, 149 (1997)
  [arXiv:hep-ph/9701278].

\bibitem{Bertolini:2000dy}
  S.~Bertolini, J.~O.~Eeg and M.~Fabbrichesi,
  Phys.\ Rev.\  D {\bf 63}, 056009 (2001)
  [arXiv:hep-ph/0002234].

\bibitem{Pallante:2001he}
  E.~Pallante, A.~Pich and I.~Scimemi,
  Nucl.\ Phys.\  B {\bf 617}, 441 (2001)
  [arXiv:hep-ph/0105011].

\bibitem{Hambye:1999yy}
  T.~Hambye, G.~O.~Kohler, E.~A.~Paschos and P.~H.~Soldan,
  Nucl.\ Phys.\  B {\bf 564}, 391 (2000)
  [arXiv:hep-ph/9906434].

\bibitem{Bosch:1999wr}
  S.~Bosch, A.~J.~Buras, M.~Gorbahn, S.~Jager, M.~Jamin, M.~E.~Lautenbacher and L.~Silvestrini,
  Nucl.\ Phys.\  B {\bf 565}, 3 (2000)
  [arXiv:hep-ph/9904408].

\bibitem{Maiani:1990ca}
  L.~Maiani and M.~Testa,
  Phys.\ Lett.\  B {\bf 245}, 585 (1990).

\bibitem{Bernard:1987pr}
  C.~W.~Bernard, T.~Draper, G.~Hockney and A.~Soni,
  Nucl.\ Phys.\ Proc.\ Suppl.\  {\bf 4}, 483 (1988).

\bibitem{Aoki:1997ev}
  S.~Aoki {\it et al.}  [JLQCD Collaboration],
  Phys.\ Rev.\  D {\bf 58}, 054503 (1998)
  [arXiv:hep-lat/9711046].

\bibitem{Dawson:1997ic}
  C.~Dawson, G.~Martinelli, G.~C.~Rossi, C.~T.~Sachrajda, S.~R.~Sharpe, M.~Talevi and M.~Testa,
  Nucl.\ Phys.\  B {\bf 514}, 313 (1998)
  [arXiv:hep-lat/9707009].

\bibitem{Golterman:1998af}
  M.~F.~L.~Golterman and K.~C.~L.~Leung,
  Phys.\ Rev.\  D {\bf 58}, 097503 (1998)
  [arXiv:hep-lat/9805032].

\bibitem{Bijnens:1984ec}
  J.~Bijnens, H.~Sonoda and M.~B.~Wise,
  Phys.\ Rev.\ Lett.\  {\bf 53}, 2367 (1984).

\bibitem{Bijnens:1984qt}
  J.~Bijnens,
  Phys.\ Lett.\  B {\bf 152}, 226 (1985).

\bibitem{Golterman:1997st}
  M.~F.~L.~Golterman and K.~C.~L.~Leung,
  Phys.\ Rev.\  D {\bf 57}, 5703 (1998)
  [arXiv:hep-lat/9711033].

\bibitem{Golterman:2000fw}
  M.~Golterman and E.~Pallante,
   ``On the determination of nonleptonic kaon decays from $K\rightarrow\pi$ matrix
  JHEP {\bf 0008}, 023 (2000)
  [arXiv:hep-lat/0006029].

\bibitem{Bijnens:1998mb}
  J.~Bijnens, E.~Pallante and J.~Prades,
  Nucl.\ Phys.\  B {\bf 521}, 305 (1998)
  [arXiv:hep-ph/9801326].

\bibitem{Cirigliano:1999pv}
  V.~Cirigliano and E.~Golowich,
  Phys.\ Lett.\  B {\bf 475}, 351 (2000)
  [arXiv:hep-ph/9912513].


\bibitem{Bertolini:1998vd}
  S.~Bertolini, M.~Fabbrichesi and J.~O.~Eeg,
  Rev.\ Mod.\ Phys.\  {\bf 72}, 65 (2000)
  [arXiv:hep-ph/9802405].

\bibitem{Buchalla:1995vs}
  G.~Buchalla, A.~J.~Buras and M.~E.~Lautenbacher,
  Rev.\ Mod.\ Phys.\  {\bf 68}, 1125 (1996)
  [arXiv:hep-ph/9512380].

\bibitem{Buras:1998raa}
  A.~J.~Buras,
  arXiv:hep-ph/9806471.

\bibitem{Ecker:1985vv}
  G.~Ecker and W.~Grimus,
  Nucl.\ Phys.\  B {\bf 258}, 328 (1985).

\bibitem{Bernard:1992mk}
  C.~W.~Bernard and M.~F.~L.~Golterman,
  Phys.\ Rev.\  D {\bf 46}, 853 (1992)
  [arXiv:hep-lat/9204007];
  C.~W.~Bernard and M.~F.~L.~Golterman,
  Phys.\ Rev.\  D {\bf 49}, 486 (1994)
  [arXiv:hep-lat/9306005].

\bibitem{Gasser:1984gg}
  J.~Gasser and H.~Leutwyler,
  Nucl.\ Phys.\  B {\bf 250}, 465 (1985).

\bibitem{Bijnens:1983ye}
  J.~Bijnens and M.~B.~Wise,
  Phys.\ Lett.\  B {\bf 137}, 245 (1984).

\bibitem{Bernard:1985wf}
  C.~W.~Bernard, T.~Draper, A.~Soni, H.~D.~Politzer and M.~B.~Wise,
  Phys.\ Rev.\  D {\bf 32}, 2343 (1985).

\bibitem{Cirigliano:2001hs}
  V.~Cirigliano and E.~Golowich,
  Phys.\ Rev.\  D {\bf 65}, 054014 (2002)
  [arXiv:hep-ph/0109265].

\bibitem{Kambor:1989tz}
  J.~Kambor, J.~Missimer and D.~Wyler,
  Nucl.\ Phys.\  B {\bf 346}, 17 (1990).

\bibitem{Crewther:1985zt}
  R.~J.~Crewther,
  Nucl.\ Phys.\  B {\bf 264}, 277 (1986).

\bibitem{Buchler:2005xn}
  M.~Buchler,
  Phys.\ Lett.\  B {\bf 633}, 497 (2006)
  [arXiv:hep-ph/0511087].

\bibitem{Blum:2001xb}
  T.~Blum {\it et al.}  [RBC Collaboration],
  Phys.\ Rev.\  D {\bf 68}, 114506 (2003)
  [arXiv:hep-lat/0110075].

\end{thebibliography}
\end{document}